\pdfoutput=1
\documentclass[camera]{jpaper}

\usepackage[sort,compress]{cite}

\newcommand{\ignore}[1]{}
\usepackage{fancyhdr}
\usepackage[normalem]{ulem}

\usepackage{mathtools}
\usepackage{titlesec}
\usepackage{microtype} 

\usepackage{array}

\makeatletter
\let\MYcaption\@makecaption
\makeatother

\usepackage{fixltx2e}

\usepackage[usenames,dvipsnames,table]{xcolor}
\usepackage{blindtext}

\usepackage{xspace}
\usepackage{ifthen}

\usepackage{booktabs}
\usepackage[all]{nowidow}
\usepackage{multirow}

\usepackage{comment}
\usepackage{enumitem}
\usepackage{setspace}
\usepackage{booktabs}
\usepackage{multirow}
\usepackage{amssymb}
\usepackage{url}

\usepackage{tikz}
\usetikzlibrary{arrows}
\usetikzlibrary{patterns}
\usetikzlibrary{calc}
\usetikzlibrary{shapes}
\usetikzlibrary{positioning,fit}
\usepackage{float}
\usepackage{lipsum}

\usepackage[nolessnomore, italic]{mathastext}
\usepackage[T1]{fontenc}
\usepackage{hhline}
\usepackage[normalem]{ulem}
\usepackage{indentfirst}
\usepackage{footmisc}

\usepackage[us,12hr]{datetime}
\usepackage{verbatim}
\usepackage{float}
\usepackage{gensymb}

\usepackage{graphicx}
\usepackage{textcomp}

\usepackage{amsmath,amssymb,amsfonts}
\usepackage{algorithm}
\usepackage[noend]{algpseudocode}
\algrenewcommand\alglinenumber[1]{\fontsize{7}{7}\selectfont #1:}

\usepackage{etoolbox}

\usepackage{stfloats}

\newcommand{\specialcellll}[2][c]{%
	\begin{tabular}[#1]{@{}c@{}}#2\end{tabular}}

\usepackage{tabularx}
\newcolumntype{Z}{>{\raggedleft\let\newline\\\arraybackslash\hspace{0pt}}X}
\newcolumntype{Y}{>{\raggedright\let\newline\\\arraybackslash\hspace{0pt}}X}

\usepackage{multirow}

\usepackage{algorithm}
\usepackage[noend]{algpseudocode}
\usepackage{etoolbox}

\makeatletter
\newcommand*{\algrule}[1][\algorithmicindent]{%
  \makebox[#1][l]{%
    \hspace*{.2em}
    \vrule height .75\baselineskip depth .25\baselineskip
  }
}

\newcount\ALG@printindent@tempcnta
\def\ALG@printindent{%
    \ifnum \theALG@nested>0
    \ifx\ALG@text\ALG@x@notext
    \else
    \unskip
    \ALG@printindent@tempcnta=1
    \loop
    \algrule[\csname ALG@ind@\the\ALG@printindent@tempcnta\endcsname]%
    \advance \ALG@printindent@tempcnta 1
    \ifnum \ALG@printindent@tempcnta<\numexpr\theALG@nested+1\relax
    \repeat
    \fi
    \fi
}
\patchcmd{\ALG@doentity}{\noindent\hskip\ALG@tlm}{\ALG@printindent}{}{\errmessage{failed to patch}}
\patchcmd{\ALG@doentity}{\item[]\nointerlineskip}{}{}{} 
\makeatother

\newcommand{\todo}[1][]{\textbf{\scriptsize \fcolorbox{black}{blue}{\color{white}{TODO}}}
\textcolor{blue}{\underline{$\overline{\hbox{\emph{#1}}}$}}}

\usepackage{tikz}
\newcommand{\circlednumber}[1]{\raisebox{0.5pt}{\protect%
  \tikz[baseline=(myanchor.base)]{
  \node[circle,fill=.,inner sep=1pt] (myanchor) {\color{-.}\footnotesize #1};}%
}}
\newcommand*\circlednumberr[1]{\raisebox{0.5pt}{\protect%
  \tikz[baseline=(char.base)]{
  \node[shape=circle,draw,densely dotted,thick,inner sep=1pt] (char) {\footnotesize #1};}%
}}

\newif\ifcameraready
\camerareadytrue

\definecolor{amber}{rgb}{1.0, 0.49, 0.0}
\definecolor{darkgreen}{rgb}{0.0, 0.2, 0.13}
\definecolor{darkbyzantium}{rgb}{0.36, 0.22, 0.33}
\definecolor{darkseagreen}{rgb}{0.56, 0.74, 0.56}
\definecolor{darkspringgreen}{rgb}{0.09, 0.45, 0.27}
\definecolor{dollarbill}{rgb}{0.52, 0.73, 0.4}

\definecolor{darkcyan}{rgb}{0.0, 0.55, 0.55}
\definecolor{forestgreen}{rgb}{0.0, 0.27, 0.13}
\definecolor{azure}{rgb}{0.0, 0.5, 1.0}
\definecolor{amber}{rgb}{1.0, 0.49, 0.0}

\definecolor{darkpink}{rgb}{0.88, 0.28, 0.54}

\newcommand{\hl}[1]{{\color{black}#1}}
\newcommand{\rev}[1]{{\color{black}#1}}
\newcommand{\sg}[1]{{\color{black}#1}}

\newcommand{\revonur}[1]{{\color{black}#1}}
\newcommand{\sgrev}[1]{{\color{black}#1}}

\newcommand{\revII}[1]{{\color{black}#1}}
\newcommand{\sgii}[1]{{\color{black}#1}}

\newcommand{\revIII}[1]{{\color{black}#1}}
\newcommand{\comm}[1]{{\color{NavyBlue}#1}}

\newcommand{\revIV}[1]{{\color{black}#1}}
\newcommand{\revV}[1]{{\color{black}#1}}

\newcommand{\revonurcan}[1]{{\color{black}#1}}

\newcommand{\revmicro}[1]{{\color{black}#1}}

 \fancypagestyle{firstpage}{
  \fancyhf{}

  \fancyfoot[C]{\thepage}
} 

\usepackage{anyfontsize}

\usepackage{graphicx,nicefrac}

\begin{document}

\setstretch{0.8}
\renewcommand{\footnotelayout}{\setstretch{0.87}}

\lineskip=0pt

\newcommand{\affilCMU}{$^{\dagger}$}
\newcommand{\affilIntel}{$^{\Join}$}
\newcommand{\affilBilkent}{$^{\triangledown}$}
\newcommand{\affilFB}{$^{\ddag}$}
\newcommand{\affilETH}{$^{\diamond}$}
\newcommand{\affilKING}{$^{\odot}$}
\newcommand{\affilUIUC}{$^{\star}$}

\title{\vspace{-20pt}GenASM: A \revonur{High-Performance, Low-Power} \\ Approximate String Matching Acceleration Framework \\ for Genome Sequence Analysis}

\author{\vspace{-24pt}\\
\protect\scalebox{0.86}{{Damla Senol Cali\affilCMU\affilIntel}\quad%
{Gurpreet S. Kalsi\affilIntel}\quad%
{Z{\"u}lal Bing{\"o}l\affilBilkent}\quad%
{Can Firtina\affilETH}\quad%
{Lavanya Subramanian\affilFB}\quad%
{Jeremie S. Kim\affilETH\affilCMU}}\\%
\protect\scalebox{0.86}{{Rachata Ausavarungnirun\affilKING}\quad%
{Mohammed Alser\affilETH}\quad%
{Juan Gomez-Luna\affilETH}\quad%
{Amirali Boroumand\affilCMU}\quad%
{Anant Nori\affilIntel}}\\%
\protect\scalebox{0.86}{{Allison Scibisz\affilCMU}\quad%
{Sreenivas Subramoney\affilIntel}\quad%
{Can Alkan\affilBilkent}\quad%
{Saugata Ghose\affilUIUC\affilCMU}\quad%
{Onur Mutlu\affilETH\affilCMU\affilBilkent}}\vspace{2pt}\\%
{\it\normalsize \affilCMU Carnegie Mellon University \quad \affilIntel \revonur{Processor Architecture Research Lab,} Intel Labs \quad \affilBilkent Bilkent University \quad \affilETH ETH Z{\"u}rich}\\
{\it\normalsize \affilFB Facebook \quad \affilKING King Mongkut's University of Technology North Bangkok \quad \affilUIUC University of Illinois at Urbana–Champaign}
\vspace{-10pt}
}

\maketitle
\ifcameraready
  \thispagestyle{plain} 
  \pagenumbering{gobble}
\else
  \thispagestyle{firstpage}
\fi
\pagestyle{plain}

\begin{abstract}

Genome sequence analysis has \revII{enabled} significant advancements in \revonur{medical and scientific} areas such as personalized medicine, outbreak tracing, and \revII{\sgii{the} understanding of} evolution. 
To perform genome sequencing, devices extract \revII{small random fragments} of an organism's DNA \revII{sequence (known as \emph{reads})}. \revII{\revIII{The} first step of genome sequence analysis is a computational process known as \emph{\revII{read mapping}}. 
In read mapping, each fragment is matched to its potential location in the reference genome with \revIII{the} goal of identifying the original location of each read in the genome.
}
Unfortunately, rapid genome sequencing is currently bottlenecked by the
computational power and memory bandwidth limitations of existing systems, as \sg{many of the steps in \revII{genome sequence analysis} must process a large amount of data. 
\revonur{A major} contributor to this bottleneck is \emph{approximate string matching} (ASM), which is used at multiple points during the \revII{mapping} process. 
ASM enables 
\revII{read mapping} to account for sequencing errors and \revIII{genetic variations} in the reads.}


We propose GenASM, \revonur{the first} ASM acceleration framework for genome sequence analysis. GenASM performs bitvector-based ASM, which can efficiently accelerate multiple steps of genome sequence analysis. 
We modify the underlying 
ASM algorithm (Bitap) to significantly increase its parallelism and reduce its memory footprint. 
\sg{Using this modified algorithm, we design the first hardware accelerator for Bitap.} \sgrev{Our hardware accelerator consists of specialized systolic-array-based compute units and on-chip SRAMs that are designed to match the rate of computation with memory capacity and bandwidth, resulting in an efficient design whose performance scales linearly as we increase the number of compute units working in parallel.} 

We demonstrate that GenASM provides significant performance and power benefits for three different use cases in genome sequence analysis.  
First, GenASM accelerates read alignment for both long reads and short reads. 
For long reads, GenASM outperforms state-of-the-art software and hardware accelerators by 116$\times$ and \revII{$3.9\times$}, respectively, while reducing power consumption by $37\times$ and 2.7$\times$.
For short reads, GenASM outperforms state-of-the-art software and hardware accelerators by $111\times$ and $1.9\times$.
Second, GenASM accelerates pre-alignment filtering for short reads, with $3.7\times$ the performance of a state-of-the-art pre-alignment filter, while \revonur{reducing power consumption by \revonur{$1.7\times$}} and significantly improving the filtering accuracy.
Third, GenASM accelerates edit distance calculation, with \revIII{22--12501$\times$} \revonur{and 9.3--400$\times$ speedups over the state-of-the-art \revonur{software} library and FPGA-based accelerator, respectively, while \revV{reducing power consumption} by \revonur{548--582$\times$} and \revIII{$67\times$}.}
\revonur{We conclude that} GenASM is a flexible, \revonur{high-performance, and low-power} framework, and we briefly discuss \revonur{four other} use cases that can benefit from GenASM.

\end{abstract}

\vspace{-5pt}
\section{Introduction} \label{sec:introduction}
\vspace{-2pt}
\revII{Genome sequencing, which determines the DNA sequence of an organism,}
\revonur{plays a pivotal role in
enabling many medical and scientific} advancements in \revonur{personalized medicine\cite{alkan2009personalized,flores2013p4,ginsburg2009genomic,chin2011cancer,Ashley2016}, evolutionary theory~\cite{ellegren2014genome,Prado-Martinez2013,Prohaska2019}, and forensics~\cite{yang2014application,borsting2015next,alvarez2017next}}.
Modern genome sequencing machines\revonur{~\cite{cali2017nanopore,minionwebpage,gridionwebpage,promethionwebpage,sequelwebpage,miseqwebpage,nextseqwebpage,novaseqwebpage}} can rapidly generate massive amounts of genomics data at low cost~\cite{Shendure2017,alser2020accelerating,mutlu2019aacbb}, but are unable to extract \revII{an} organism's \revII{complete} DNA in one piece.  Instead, these machines \revII{extract} 
\revII{smaller random} fragments of the original DNA sequence, known as \emph{reads}.
These reads \revII{then} pass through a computational process known as \emph{\revII{read mapping}}, which takes \revII{each read, aligns it to one or more possible locations within the reference genome, and finds the matches and differences (i.e., \emph{distance}) between the read and the reference genome segment at that location\revIII{~\cite{alkan2009personalized,Xin2013}}. Read mapping is the first key step in genome sequence analysis.}
 

\sg{State-of-the-art sequencing machines produce \revII{broadly} one of two kinds of reads.
\emph{Short reads} (consisting of no more than a few hundred DNA base \revonur{pairs~\cite{chaisson2004fragment,trapnell2009map}}) are generated using \revII{short-read sequencing (SRS)} technologies\revonur{~\cite{reuter2015high,van2014ten}, which have been on the market for more than a decade}. Because each read fragment is so short compared to the entire DNA (e.g., a \sgrev{human's} DNA consists of over 3~billion base \revonur{pairs}\revII{~\cite{venter2001sequence}}), short reads incur a number of reproducibility (e.g., non-deterministic mapping) and computational challenges\revIII{~\cite{Firtina2016,Xin2013,Xin2015,alkan2011limitations,treangen2011repetitive,alser2020technology,gatekeeper,xin2016optimal,mutlu2019aacbb}}.
\emph{Long reads} (\revII{consisting of} \revonur{\revIII{thousands to} millions of DNA base pairs}) are generated using \revII{long-read sequencing (LRS) technologies, of which Oxford Nanopore Technologies’ (ONT) nanopore sequencing~\cite{cali2017nanopore,lu2016oxford,magi2017nanopore,clarke2009continuous,deamer2016three,marx2015nanopores,branton2008potential,laver2015assessing,ip2015minion,kasianowicz1996characterization,jain2018nanopore,quick2014reference} and Pacific Biosciences’ (PacBio) single-molecule real-time (SMRT) sequencing~\cite{english2012mind,roberts2013advantages,rhoads2015pacbio,wenger2019accurate,nakano2017advantages,van2018third,mantere2019long,amarasinghe2020opportunities}} are the most \revII{widely used ones}. 
\revIII{LRS} technologies are relatively \revII{new}, and they avoid} many of the challenges faced by short reads.

\revII{LRS technologies} have \revII{three} key advantages compared to \revII{SRS technologies}.
\revII{First, LRS devices can generate very long reads, which 
\sgii{(1)~reduces the non-deterministic mapping problem faced by short reads, as long reads are significantly more likely to be unique
and therefore have fewer potential mapping locations in the reference genome; and
(2)~span larger parts of the \revIII{repeated or} complex regions of a genome, enabling detection of \revIII{genetic variations} \revIV{that might} exist in these regions~\cite{van2018third}.}
Second, \sgii{LRS \revIII{devices
perform} real-time sequencing, and can enable concurrent sequencing and analysis~\cite{quick2016real,roberts2013advantages,logsdon2020long}.}
Third, ONT's pocket-sized device (MinION~\cite{minionwebpage}) provides portability, making sequencing possible at remote places using laptops or mobile devices.}
This enables a number of new applications, 
such as rapid infection diagnosis and outbreak tracing (e.g., COVID-19, Ebola, Zika, \revIII{swine flu\revonur{~\cite{quick2016real,wu2020new,harcourt2020isolation,james2020lampore,da2020evolution,greninger2015rapid,wang2015minion,faria2016mobile}}).
Unfortunately,} \revII{LRS} devices are much more error-prone in sequencing (with a typical error rate of 10--15\%~\cite{jain2018nanopore, weirather2017comprehensive,ardui2018single,van2018third}) compared to \revII{SRS} \revonur{devices} (typically 0.1\%~\cite{glenn2011field,quail2012tale,goodwin2016coming}), \revonur{which leads to new computational challenges\revIII{~\cite{cali2017nanopore}}.}

For both short and long reads, \emph{multiple} steps of \revII{read mapping} must account for the \revonur{sequencing errors, \revII{and \sgii{for} the differences} caused by genetic mutations and variations}. These \revIII{errors and} \revII{differences} take the form of base insertions, deletions, and/or substitutions\revonur{~\cite{navarro2001guided,waterman1976some,smith1981identification,wu1992fast,myers1999fast,ukkonen1985algorithms}}.
As a result, \revII{read mapping} must perform \emph{approximate} (or \emph{fuzzy}) \emph{string matching} (ASM). Several algorithms exist for ASM, but state-of-the-art \revII{read mapping} tools typically make use of an expensive dynamic programming based algorithm\revonur{~\cite{smith1981identification,levenshtein1966binary,needleman1970general}} that scales quadratically in both execution time and required storage. This ASM algorithm has been shown to be the major bottleneck in \revII{read mapping}~\cite{gatekeeper,turakhia2018darwin,fujiki2018genax,alser2020accelerating,ham2020genesis,nag2019gencache,huangfu2019medal}. Unfortunately, as sequencing technologies advance, the growth in the rate that sequencing devices generate reads is far outpacing the corresponding growth in computational power~\cite{check2014technology, alser2020accelerating}, placing greater pressure on the ASM bottleneck.
Beyond \revII{read mapping}, ASM is a key technique for \revII{other bioinformatics problems such as whole genome alignment (WGA)~\cite{delcher1999alignment,kurtz2004versatile,bray2003avid,hohl2002efficient,schwartz2003human,brudno2003lagan,dewey2019whole,darwin-wga,lin2020gsalign,marccais2018mummer4,li2018minimap2} and multiple sequence alignment (MSA)~\cite{sankoff1975minimal,carrillo1988multiple,paten2011cactus,higgins1988clustal,lipman1989tool,notredame2000t,lee2002multiple,notredame2002recent,edgar2006multiple}, where two or more whole genomes, or regions of multiple genomes (from the same or different species), are compared to determine their similarity 
for predicting evolutionary relationships or finding common regions (e.g., genes).}
Thus, there is a \revII{pressing} need to develop \revII{techniques} for \revII{genome sequence analysis} that \sgrev{provide} \revonur{fast and efficient} ASM.

In this work, we propose \emph{GenASM}, \sg{an ASM} \revIV{acceleration} framework for \revII{genome sequence analysis}. Our goal is to design a fast, efficient, and flexible framework for both short and long reads, which \sg{can be used to accelerate} \emph{multiple steps} of the genome sequence analysis pipeline. 
\sg{To avoid implementing more complex hardware for the dynamic programming \revII{based} algorithm\revonur{~\cite{Fei2018, kaplan2019poster, turakhia2018darwin, gupta2019rapid, Banerjee2019,jiang2007reconfigurable,rucci2018swifold, chen2014accelerating}, we base GenASM upon}
the \textit{Bitap} algorithm~\cite{baeza1992new, wu1992fast}.
Bitap uses only fast and simple bitwise operations to perform 
\revonur{approximate} string matching, making it amenable to \revonur{efficient} hardware acceleration. To our knowledge, GenASM is the first work that enhances and accelerates Bitap.}


\revIV{To} use Bitap for GenASM, we make two key \revonur{algorithmic} modifications that allow us to overcome key limitations that \revonur{prevent the original Bitap} algorithm from being efficient for genome \revII{sequence analysis} (we discuss these limitations in Section~\ref{sec:background-bitap}).
First, to improve Bitap's \revonur{applicability to different sequencing technologies and its performance, we (1)~modify the algorithm to support long reads \revII{(in addition to already supported short reads)}, and (2)~eliminate loop-carried data dependencies so that we can parallelize a single string matching operation. Second, we develop a \revIV{novel} Bitap-compatible algorithm for \revII{\emph{traceback}}, 
\revIII{a method that} utilizes information collected during ASM about the different types of errors
to identify the optimal alignment \sg{of reads}. \revII{The original Bitap algorithm is not capable of performing traceback.}}



In GenASM, we \emph{co-design} our modified Bitap \sgii{algorithm 
and} our new Bitap-compatible \revII{\emph{traceback}}
algorithm with an area- and power-efficient hardware \sgrev{accelerator, which consists} of two components:
(1)~\emph{GenASM-DC}, which provides hardware support to efficiently execute our modified Bitap algorithm to generate bitvectors \revII{(each \sgii{of which} represents \revIII{one of the four possible cases: match, insertion, deletion, or substitution)}}
and perform distance calculation~(DC) \revII{(\sgii{which calculates the} minimum number of errors between the read and the reference segment)}; and
(2)~\emph{GenASM-TB}, which provides hardware support to efficiently execute our novel traceback~(TB) algorithm to find the optimal alignment \sgrev{of a read,} using the bitvectors generated by GenASM-DC.
Our hardware \sgrev{accelerator} \revII{(1)~\sgrev{balances} the compute resources \sgii{with} available memory capacity \revII{and bandwidth} per compute unit \sgrev{to avoid wasting resources}, (2)~\sgrev{achieves} high performance and \revII{power} efficiency \sgrev{by using specialized compute units that we design to exploit} data locality, and (3)~\sgrev{scales linearly in performance with the number of parallel compute units that we add to the system}.}

\textbf{Use Cases.} GenASM is an efficient framework for accelerating genome sequence analysis \sg{that has multiple possible use cases}. In this paper, we describe and \revonur{rigorously} evaluate three use cases of GenASM. First, we show that GenASM \revonur{can effectively accelerate} the read alignment 
step of read \revII{mapping} \revV{(Section~\ref{sec:results-alignment})}.  
Second, we illustrate \revonur{that GenASM can be employed as the most efficient (to date)} pre-alignment filter\revIII{~\cite{gatekeeper,Alser2019}} for short reads
\revV{(Section~\ref{sec:results-filtering})}. Third, we demonstrate how GenASM can \revonur{efficiently find the edit distance (i.e., Levenshtein distance~\cite{levenshtein1966binary})} between two sequences of arbitrary lengths \revV{(Section~\ref{sec:results-edit})}. \sg{In addition, GenASM can be utilized in several other parts of \revII{genome} sequence analysis \revIII{as well as \revIV{in} text analysis}, which we briefly discuss} in Section~\ref{sec:bitmac-framework-other}.


\textbf{Results Summary.} We evaluate GenASM for three different use cases of ASM in genome sequence analysis using a combination of \revonur{the synthesized SystemVerilog model of our hardware accelerators and detailed simulation-based performance modeling}. (1)~For read alignment, we compare GenASM to state-of-the-art software (Minimap2~\cite{li2018minimap2} and BWA-MEM~\cite{li2013aligning}) and hardware approaches (GACT in Darwin~\cite{turakhia2018darwin} and SillaX in GenAx~\cite{fujiki2018genax}), and find that GenASM is significantly more efficient in terms of both speed and power consumption. For this use case, we compare GenASM \emph{only} with the read alignment 
steps of the baseline tools and accelerators.
For long reads, GenASM achieves $116\times$ and $648\times$ speedup over 12-thread runs of the alignment steps of Minimap2 and BWA-MEM, respectively, while reducing power consumption by $37\times$ and $34\times$.
Compared to GACT, GenASM provides \revII{$6.6\times$ the throughput 
per unit area and $10.5\times$ the throughput per unit power} 
for long reads.
For short reads, GenASM achieves $158\times$ and $111\times$ speedup over 12-thread runs of the \revIII{alignment} steps of Minimap2 and BWA-MEM, respectively, while reducing power consumption by $31\times$ and $33\times$.
Compared to SillaX, GenASM \revII{is} 1.9$\times$ \revII{faster} at a comparable area and power consumption.
(2)~For pre-alignment filtering of short reads, we compare GenASM with \revII{a} state-of-the-art FPGA-based filter, Shouji~\cite{Alser2019}. GenASM provides $3.7\times$ speedup over Shouji, while \revonur{reducing power consumption by \revonur{$1.7\times$}}, and \revII{also} significantly improving the filtering accuracy. (3)~For edit distance calculation, we compare GenASM with \revII{a} state-of-the-art software library, Edlib~\cite{vsovsic2017edlib}\revII{, and} \revonur{FPGA-based accelerator, ASAP~\cite{Banerjee2019}}. Compared to Edlib, GenASM \revIV{provides} \revIII{22--12501$\times$ speedup, for varying sequence lengths and similarity values,} \revonur{while reducing power consumption by 548--582$\times$. Compared to ASAP, GenASM \revIV{provides} 9.3--400$\times$ speedup, while reducing power consumption by \revIII{$67\times$}.}



This paper makes the following contributions:
\begin{itemize}[itemsep=0pt, topsep=0pt, parsep=0pt, leftmargin=*]

\item To our knowledge, GenASM is the \emph{first} work that enhances and accelerates the Bitap algorithm for approximate string matching. We modify Bitap to add efficient support for long reads and enable parallelism \revII{within each ASM operation}. We also propose the \emph{first} Bitap-compatible traceback algorithm. \revIV{We open source our software implementations of \revV{the GenASM 
algorithms}~\cite{genasmgithub}}.

\item We present GenASM, a novel approximate string matching acceleration framework for genome sequence analysis. GenASM is a power- and area-efficient hardware implementation of our new Bitap-based algorithms. 

\item We show that GenASM can accelerate \emph{three use cases} of approximate string matching (ASM) in genome sequence analysis (\revV{i.e.,} read alignment, pre-alignment filtering, edit distance \revIII{calculation).
We} \sg{find that GenASM is \revonur{greatly faster and more \revII{power-}efficient} for all three use cases than state-of-the-art software and hardware baselines.}


\end{itemize}

\vspace{-5pt}
\section{Background} \label{sec:background}
\vspace{-2pt}

\vspace{-2pt}
\subsection{Genome Sequence Analysis Pipeline}\label{sec:background:pipeline}
\vspace{-2pt}

\begin{figure}[h!]
\centering
\vspace{-4pt}
\includegraphics[width=7.5cm,keepaspectratio]{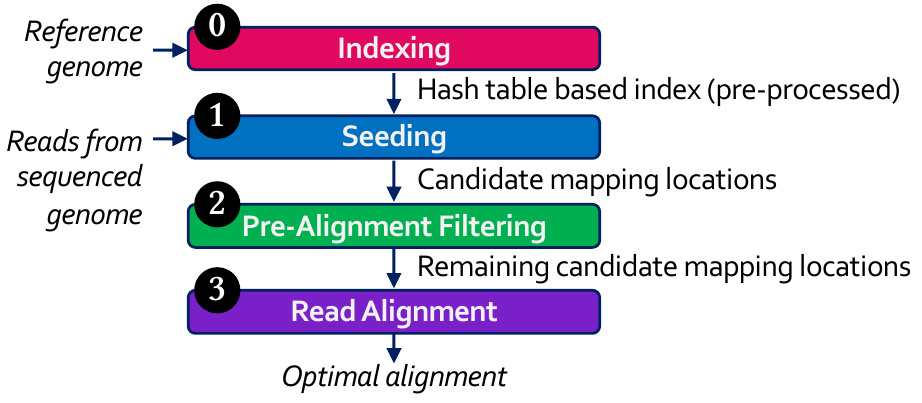}
\vspace{-5pt}
\caption{Four steps of read mapping.} 
\label{fig:pipeline}
\vspace{-4pt}
\end{figure}

A common approach to \revonur{the first step in genome sequence analysis} is to perform \revII{\textit{read mapping}, where \sgii{each \textit{read} of an organism's sequenced} genome \sgii{is} matched against the \textit{reference genome for the organism's species} to find \sgii{the read's original location.}}
As Figure~\ref{fig:pipeline} shows, \revonur{typical}
read mapping\revonur{~\cite{li2018minimap2,li2013aligning,alkan2009personalized,Xin2013,langmead2012fast,li2009soap2}} is a four-step process. 
First, read mapping starts with \textit{indexing}~\circlednumber{0}, \revII{which is an offline pre-processing step performed on a known reference genome}. Second, \sgrev{once a sequencing machine generates reads from a DNA sequence, the} \textit{seeding} \sgrev{process}~\circlednumber{1} queries the index structure \sgii{to determine} \revII{the candidate \revIII{(i.e., potential)} mapping locations} of each read 
in the reference genome using substrings (i.e., \textit{seeds}) from each read. Third, \sgii{for each read,} \textit{pre-alignment filtering}~\circlednumber{2} uses filtering heuristics to examine the similarity between \sgii{a} read and 
\sgii{the portion of the reference genome at each of the read's candidate mapping locations.}
These filtering heuristics aim to eliminate most of the dissimilar \revII{pairs of reads and candidate \sgii{mapping locations}} to decrease the number of required alignments \revIII{in the next step}. Fourth, for all of the \sgii{remaining} candidate \revII{mapping locations}, \textit{read alignment}~\circlednumber{3}
\revII{runs a dynamic programming based algorithm} to determine 
\sgii{which of the candidate mapping locations in the reference matches best with the input read.}
As part of this step, traceback \sgii{is performed} between the reference 
and the \revonur{input \sgii{read to} find the \emph{optimal alignment}, which is the alignment with the highest \sgrev{likelihood of being correct (based on a scoring \revII{function}\revIII{~\cite{gotoh1982improved,miller1988sequence,waterman1984efficient}})}}.
\revonur{\sgrev{The optimal} alignment is defined using a \emph{CIGAR string}\revII{~\cite{li2009sequence}}, which shows the sequence \sgrev{and position of each match, substitution, insertion, and deletion for the read \revII{with respect to the \sgii{selected mapping location}} of the reference.}}

\subsection{Approximate String Matching (ASM)}
\label{sec:approx-background}
\vspace{-2pt}

\begin{figure}[t!]
\centering
\vspace{-1pt}
\includegraphics[width=6cm,keepaspectratio]{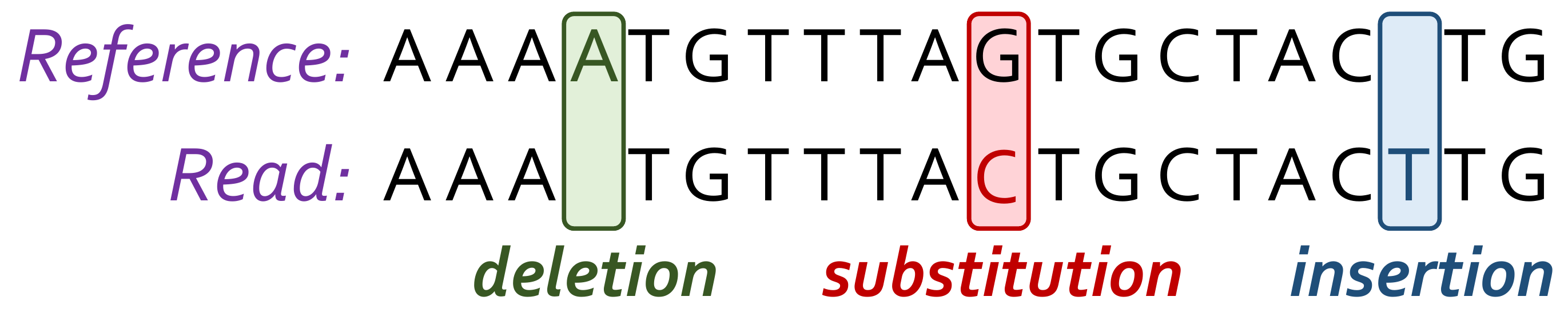}
\vspace{-4pt}
\caption{Three types of errors (i.e., edits).}\label{fig:edits}
\vspace{-6pt}
\end{figure}

The goal of approximate string matching \cite{navarro2001guided} is to detect the differences and similarities between \revII{two sequences. \revIII{Given a query read sequence~$Q$=[$q_1$$q_2$\ldots$q_m$], a reference text sequence~$T$=[$t_1$$t_2$\ldots$t_n$] (where $m=|Q|$, $n=|T|$, $n \geq m$),}} and an edit distance threshold $E$, the approximate string matching problem is to identify a set of approximate matches of \revII{$Q$ in $T$} (allowing for at most $E$ differences). 
\revonur{The differences between two sequences of the same species \sgrev{can result from sequencing errors\revIII{~\cite{fox2014accuracy,amarasinghe2020opportunities}} and/or genetic variations\revIII{~\cite{feuk2006structural,alkan2011genome}}}}.
\revII{Reads are prone to sequencing errors, \sgii{which account for about 0.1\% of the length of short reads\revIII{~\cite{glenn2011field,quail2012tale,goodwin2016coming}} and 
\revIII{10--15\%} of the length of} long reads\revIII{~\cite{jain2018nanopore, weirather2017comprehensive,ardui2018single,van2018third}}.


The differences, known as \emph{edits},} can be classified as \textit{substitutions}, \textit{deletions}, \sgrev{or} \textit{insertions} in one or both sequences~\cite{levenshtein1966binary}. Figure~\ref{fig:edits} shows each possible kind of edit. \revII{In ASM, to detect} a deleted character or an inserted \revIII{character, 
we} need to examine all possible \sgrev{\emph{prefixes}} (i.e., substrings that include the first character of the string) \revIII{or \revV{\emph{suffixes}} (i.e., substrings that include the last character of the string)} of the two input sequences, and keep track of the pairs of prefixes \revIII{or suffixes} that provide the \revonur{minimum number of edits}.

Approximate string matching is needed not only to determine the minimum number of edits between two genomic sequences, but also to provide the location and type of each edit. As \revII{two} sequences could have a large number of different possible arrangements of the edit operations and matches (and hence different \revII{\emph{alignments}}), the approximate string matching algorithm usually involves a traceback step. 
\revII{The alignment score is the sum of all edit penalties and match scores along the alignment, as defined by a user-specified scoring function.}
This step finds the \revII{\emph{optimal alignment} as the combination of edit operations to build up the highest alignment score.}

Approximate string matching is typically implemented as a dynamic programming 
\revII{based} algorithm. Existing implementations, such as Levenshtein distance \cite{levenshtein1966binary}, Smith-Waterman \cite{smith1981identification}, and Needleman-Wunsch \cite{needleman1970general}, have quadratic time and space complexity (i.e., $O(m \times n)$ between two sequences with lengths $m$ and $n$). \revII{Therefore, it is desirable to find lower-complexity algorithms for ASM.}

\vspace{-1pt}
\subsection{Bitap Algorithm}
\label{sec:background-bitap}
\vspace{-2pt}

One candidate to replace dynamic programming \revII{based} \revIII{algorithms} for \revIV{ASM} is the \textit{Bitap} algorithm~\cite{baeza1992new, wu1992fast}.
Bitap tackles the problem of computing the minimum edit distance between a \revonur{reference} text (e.g., reference genome) and a query pattern (e.g., read) with a maximum of \textit{k} many errors. When \textit{k} is 0, the algorithm finds the exact matches. 

\begin{figure}[t!]
\centering
\vspace{-3pt}
\includegraphics[width=\columnwidth,keepaspectratio]{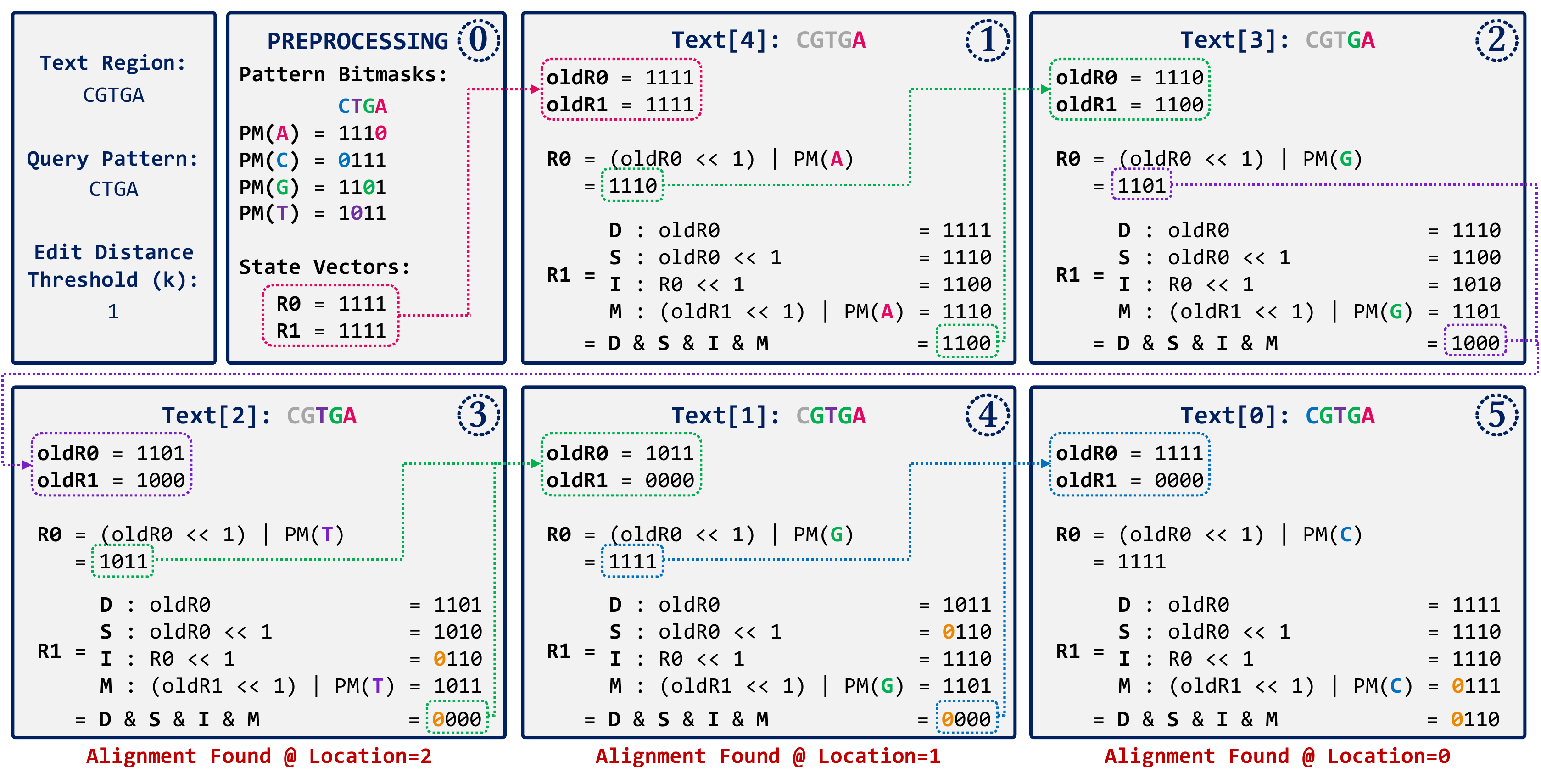}
\vspace{-16.5pt}
\caption{\revonur{Example for the Bitap algorithm.}} \label{fig:bitap-dc-example}
\vspace{-6pt}
\end{figure}

\revonur{
Algorithm~\ref{bitap-search-alg} shows the \textit{Bitap} algorithm and Figure~\ref{fig:bitap-dc-example} shows an example for the execution of the algorithm. The algorithm starts with a pre-processing procedure (\revIII{Line~4} in Algorithm~\ref{bitap-search-alg}; \circlednumberr{0} in Figure~\ref{fig:bitap-dc-example}) that converts the query pattern into $m$-sized pattern bitmasks, \textit{PM}. 
We generate one pattern bitmask for each character in the alphabet. \revII{Since $0$ means match in the Bitap algorithm, we} \revII{set} $PM[a][i]=0$ \revIII{when} $pattern[i] = a$, where $a$ is a character from the alphabet (\revIII{e.g.,} A, C, G, T). 
These pattern bitmasks help us to represent the \revIII{query pattern} in a binary format.
After the bitmasks are prepared for each character, \sgii{every bit} of all \revII{status \sgii{bitvectors} ($R[d]$, where $d$ is in range $[0,k]$)} \sgii{is initialized to 1} (\revIII{Lines~5--6} in Algorithm~\ref{bitap-search-alg}; \circlednumberr{0} in Figure~\ref{fig:bitap-dc-example}). \revII{Each $R[d]$ bitvector at text iteration $i$ holds the partial match information between $text[i:(n-1)]$ \revV{(Line~8)} and the query with maximum of $d$ \revIII{errors. 
Since} at the beginning of the execution there are no matches, we initialize all status bitvectors with 1s.}
The \revII{status} \sgii{bitvectors} of the previous iteration with edit distance $d$ is kept in $oldR[d]$ 
(\revIII{Lines~10--11}) 
to take \revII{partial matches into consideration in the next iterations}.}

\revII{The algorithm examines each text character one by one, one per iteration.} At each text iteration (\circlednumberr{1}--\circlednumberr{5}), the \revII{pattern} bitmask of the current text character ($PM$) is retrieved \revIII{(Line~12)}. \revV{After the status bitvector for exact match is computed ($R[0]$; Line~13), the}
\sgii{status bitvectors for each distance \revV{($R[d]; d=1...k$)} are computed using the rules in \revIII{Lines~15--19}.
For a distance $d$, three intermediate bitvectors for the error cases (one each for deletion, insertion, substitution; D/I/S in Figure~\ref{fig:bitap-dc-example}) are calculated by using $oldR[d-1]$ or $R[d-1]$, since a new error is being added (i.e., the distance is increasing by 1), while the intermediate bitvector \revV{for the} match case (M) is calculated using $oldR[d]$.}
\sgii{For a deletion \revIII{(Line~15)}, we are looking for \revIII{a string match} if the current pattern character is missing, so we copy the partial match information of the previous character ($oldR[d-1]$; consuming a text character) \emph{without} any shifting (\emph{not} consuming a pattern character) to serve as the deletion bitvector \revIII{(labeled as $D$ of $R1$ bitvectors in \circlednumberr{1}--\circlednumberr{5})}.
For a substitution \revIII{(Line~16)}, we are looking for \revIII{a string match} if the current pattern character and \revV{the} current text character do not match, so we take the partial match information of the previous character ($oldR[d-1]$; consuming a text character) and shift it left by one (consuming a pattern character) before saving it as the substitution bitvector \revIII{(labeled as $S$ of $R1$ bitvectors in \circlednumberr{1}--\circlednumberr{5})}.
For an insertion \revIII{(Line~17)}, we are looking for \revIII{a string match} if the current text character is missing, so we copy the partial match information of the \emph{current} character ($R[d-1]$; \emph{not} consuming a text character) and shift it left by one (consuming a pattern character) before saving it as the insertion bitvector \revIII{(labeled as $I$ of $R1$ bitvectors in \circlednumberr{1}--\circlednumberr{5})}.
For a match \revIII{(Line~18)}, we are looking for \revIII{a string match} only if the current pattern character matches the current text character, so we take the partial match information of the previous character ($oldR[d]$; consuming a text character but \emph{not} increasing the edit distance), shift it left by one (consuming a pattern character), \revII{and perform an OR operation with the pattern bitmask of the current text character ($curPM$; comparing the text character and the pattern character)} before saving the result as the match bitvector \revIII{(labeled as $R0$ bitvectors and $M$ of $R1$ bitvectors in \circlednumberr{1}--\circlednumberr{5})}.}

{
\begin{algorithm}[t!]
\caption{Bitap Algorithm}\label{bitap-search-alg}
\fontsize{7}{7}\selectfont{\revIII{\textbf{Inputs:} \texttt{text} (reference), \texttt{pattern} (query), \texttt{k} (edit distance threshold)\\
\textbf{Outputs:} \texttt{startLoc} (matching location), \texttt{editDist} (minimum edit distance)}}
\vspace{-1pt}
\begin{algorithmic}[1]
\fontsize{7}{7}\selectfont
    \State $\texttt{n} \gets \texttt{length of \revIII{reference text}}$
    \State $\texttt{m} \gets \texttt{length of \revIII{query pattern}}$
    \Procedure{Pre-Processing}{}
        \State $\texttt{PM} \gets $\texttt{generatePatternBitmaskACGT(pattern)}
        \Comment{\comm{pre-process the pattern}}
        \For{\texttt{d in 0:k }}
            \State $\texttt{R[d]} \gets \texttt{111..111}$ \Comment{\comm{initialize R bitvectors to 1s}}
        \EndFor
    \EndProcedure
    \Procedure{Edit Distance Calculation}{}
        \For{\texttt{i in (n-1):\revV{-1:}0}}
        \Comment{\comm{iterate over each text character}}
            \State $\texttt{curChar} \gets \texttt{text[i]}$
            \For{\texttt{d in 0:k }}
                \State $\texttt{oldR[d]} \gets \texttt{R[d]}$ \Comment{\comm{copy previous iterations' bitvectors as oldR}}
            \EndFor
            \State $\texttt{curPM} \gets \texttt{PM[curChar]}$
            \Comment{\comm{retrieve the pattern bitmask}}
            \State $\texttt{R[0]} \gets \texttt{(oldR[0]}\verb|<<|1)\texttt{ | curPM}$ \Comment{\comm{status bitvector for exact match}}
            \For{\texttt{d in 1:k }}
            \Comment{\comm{iterate over each edit distance}}
                \State $\texttt{deletion \revIII{(D)}} \gets \texttt{oldR[d-1]}$
                \State $\texttt{substitution \revIII{(S)}} \gets \texttt{(oldR[d-1]}\verb|<<|\texttt{1)}$
                \State $\texttt{insertion \revIII{(I)}} \gets \texttt{(R[d-1]}\verb|<<|\texttt{1)}$
                \State $\texttt{match \revIII{(M)}} \gets \texttt{(oldR[d]}\verb|<<|\texttt{1)}\texttt{ | curPM}$
                \State $\texttt{R[d]} \gets \texttt{\revIII{D \& S \& I \& M}}$
                \Comment{\comm{status bitvector for $d$ errors}}
            \EndFor
            \If {\texttt{MSB of R[d] == 0, where 0 $\leq$ d $\leq$ k}}
            \Comment{\comm{check if MSB is 0}}
                \State $\texttt{startLoc} \gets \texttt{i}$
                \Comment{\comm{matching location}}
                \State $\texttt{editDist} \gets \texttt{d}$
                \Comment{\comm{found minimum edit distance}}
            \EndIf
        \EndFor
    \EndProcedure
\end{algorithmic}
\end{algorithm}
}

\revII{After computing all four \sgii{intermediate} bitvectors,
in order to take all possible partial matches into consideration, we perform an \sgii{AND operation} \revIII{(Line~19)} with these four bitvectors to \revIII{preserve} all 0s \revIII{that exist in any of them 
(i.e., all potential locations for a string match with 
an edit distance of $d$ up to this point)}. \sgii{We save the ANDed result as the} $R[d]$ status bitvector for the current iteration. \sgii{This process is repeated for each potential edit distance value from 0 to $k$.}}
\sgii{If} \revII{the most significant bit of \sgii{the} $R[d]$ bitvector becomes 0 \revIII{(Lines~20--22)}, \revIV{then there} is a match starting at position $i$ of the text with an edit distance $d$ \revIII{(as shown in \circlednumberr{3}--\circlednumberr{5}).}}
The traversal of the text \sgii{then} continues until all possible text positions are \sgii{examined.}

\vspace{-5pt}
\section{Motivation and Goals} \label{sec:motivation}
\vspace{-3pt}

\sgrev{Although the Bitap algorithm is highly suitable for hardware acceleration due to the simple nature of its bitwise operations, \revIV{we find that} it has five limitations that hinder its \revIII{applicability and} efficient hardware acceleration for \revIII{genome analysis}. In this section, we discuss each of these limitations. In order to overcome these limitations and design an effective and efficient accelerator, we find that we need to both \revIII{(1)~modify and extend} the Bitap algorithm and \revIII{(2)~develop} specialized hardware that can exploit the new opportunities that our algorithmic modifications provide.}

\revonur{

\vspace{-1pt}
\subsection{Limitations of Bitap on Existing Systems}
\label{sec:motivation:limitations}
\vspace{-3pt}

\textbf{\revII{No Support for Long Reads.}}
\sgrev{In state-of-the-art implementations of Bitap}, the query length is limited by the word size of the machine running \sgrev{the algorithm.  This is due to (1)~the fact that the bitvector length must be equal to the query length, and (2)~the need to perform bitwise operations on the bitvectors. By limiting the bitvector length to a word, each bitwise operation can be done using a single CPU instruction.
Unfortunately, the lack of multi-word queries prevents these implementations from working for long reads, whose lengths are on the order of thousands \revIII{to millions} of base pairs
(which require thousands of \revII{bits} 
to store).}

\textbf{Data Dependency Between Iterations.} 
As we show in Section~\ref{sec:background-bitap}, the computed bitvectors at each \revIII{text iteration} (i.e., R[d]) \revIII{of the Bitap algorithm} depend on the bitvectors computed in the previous \revIII{text} iteration (i.e., oldR[d-1] and oldR[d]; \revIII{Lines~11, 13, 15, 16, and 18 of Algorithm~\ref{bitap-search-alg})}. Furthermore, for each \revIII{text} character, there is an inner loop that iterates for the maximum edit distance number of iterations \revIII{(Line~14)}. The bitvectors computed in each of these inner iterations (i.e., R[d]) are also dependent on the previous inner iteration's computed bitvectors (i.e., R[d-1]; \revIII{Line~17})}. \sgrev{This two-level data dependency forces the \revII{consecutive} iterations to take place sequentially.}

\textbf{\revII{No Support for Traceback.}}
\revII{Although the baseline Bitap algorithm can find possible matching locations of each query read within the reference text, this covers only the first step of approximate string matching required for genome sequence analysis. Since there could be multiple different alignments between the read and the reference, \revIII{the} traceback \revIII{operation~\cite{myers1988optimal,gotoh1986alignment,gotoh1982improved,miller1988sequence,waterman1984efficient,altschul1986optimal,fickett1984fast,smith1981identification,waterman1976some,ukkonen1985algorithms}} is needed to find the \emph{optimal alignment}, which is the alignment with the minimum edit distance (or with the highest score based on a user-defined scoring function). However, Bitap does not include any such support for optimal alignment identification.}

\textbf{Limited Compute Parallelism.} 
\revII{Even \revIII{after} we solve the algorithmic limitations of Bitap, we find that we cannot extract significant performance benefits with just algorithmic enhancements alone. For example,}
\sgrev{while Bitap iterates over each character of the input text sequentially \revIII{(Line~8)},} we can enable \emph{text-level parallelism} \sgrev{to improve its performance (Section~\ref{sec:bitap-search}).} 
However, the \sgrev{achievable} level of parallelism is limited by the number of compute units in existing systems. For example, our studies show that Bitap is bottlenecked by computation on CPUs, since the working set fits within the private caches but the limited number of cores prevents the further speedup of the algorithm.

\textbf{Limited Memory Bandwidth.} 
\sgrev{  
We would expect that a GPU, which has thousands of compute units, can overcome the limited compute parallelism issues that CPUs experience.} 
However, \sgrev{we find that a GPU implementation of the \revIII{Bitap} algorithm suffers from the limited amount of memory bandwidth available for each GPU thread. Even} when we run a CUDA implementation of the baseline Bitap algorithm~\cite{li2011fast}, \sgrev{whose bandwidth requirements are significantly lower than our modified algorithm, the limited \revIII{memory} bandwidth bottlenecks the algorithm's performance.}
\sgrev{We find that the bottleneck is exacerbated after the number of threads per block reaches 32, as Bitap} becomes shared cache-bound (i.e., on-GPU L2 cache-bound). The small number of registers becomes insufficient to hold the intermediate data required for Bitap execution. Furthermore, when the working set of a thread does not fit within the private memory of the thread, destructive interference between threads while accessing the shared memory creates bottlenecks in the algorithm on GPUs.
\sgrev{We expect these issues to worsen when we implement traceback, which requires significantly higher bandwidth than Bitap.}

\vspace{-1pt}
\subsection{Our Goal}
\label{sec:motivation:genasm}
\vspace{-1pt}

\revII{Our goal in this work is to overcome these limitations and use Bitap in a fast, efficient, and flexible \revII{ASM} framework for both short and long reads. We find that this \revIII{goal} cannot be achieved by modifying only the algorithm or only the hardware. \revIII{We design} \emph{GenASM}, the first ASM \revIV{acceleration} framework for \revII{genome sequence analysis}. Through careful modification and co-design of the enhanced Bitap algorithm and hardware, \revIII{GenASM aims to} successfully replace the expensive dynamic programming \revII{based} algorithm used for ASM in genomics with the efficient bitwise-operation-based Bitap algorithm, which can accelerate \emph{multiple steps} of genome sequence analysis.}

\vspace{-5pt}
\section{G\lowercase{en}ASM: A High-Level Overview}
\label{sec:bitmac-overall}

\vspace{-2pt}

\begin{figure}[b!]
\centering
\vspace{-3pt}
\includegraphics[width=\columnwidth,keepaspectratio]{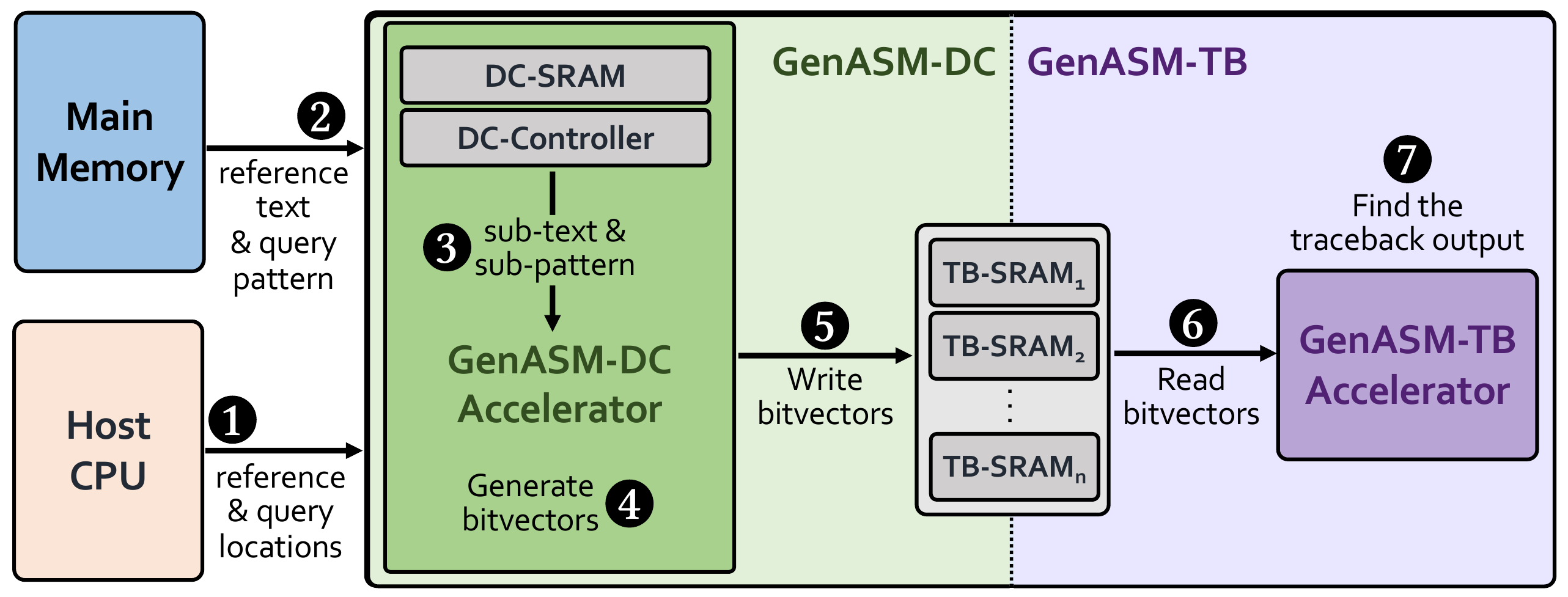}
\vspace{-16pt}
\caption{\revII{Overview of GenASM.}} \label{fig:bitmac-pipeline}
\vspace{-3pt}
\end{figure}

\revonur{In GenASM, we \emph{co-design} our modified Bitap algorithm for distance calculation (DC) and our new Bitap-compatible traceback (TB) algorithm with an area- and power-efficient hardware accelerator.
GenASM consists of two components, as shown in Figure~\ref{fig:bitmac-pipeline}:
(1)~GenASM-DC (Section~\ref{sec:bitap-search}), which for each read generates the bitvectors
and \revonur{performs the minimum edit distance calculation (DC)}; and
(2)~GenASM-TB (Section~\ref{sec:bitap-traceback}), which uses the bitvectors to perform traceback \revonur{(TB)} and find the optimal alignment. GenASM is a flexible framework \revII{that} can be used for different use cases \revIII{(Section~\ref{sec:bitmac-framework}}).}

\sgii{GenASM execution starts when} the host CPU issues \sgii{a} task to GenASM with the reference and the query sequences' locations (\circlednumber{1} in Figure~\ref{fig:bitmac-pipeline}).
GenASM-\hl{DC reads the corresponding reference \revV{text} region and the query \revV{pattern} from the memory.  GenASM-DC then writes these to its dedicated SRAM, which we call DC-SRAM (\circlednumber{2}). After that, GenASM-DC divides the reference \revIII{text (e.g., reference genome) and query pattern (e.g., read)} into multiple overlapping windows \revIII{(\circlednumber{3})}, and for each \emph{sub-text} (i.e., the portion of the reference \revIII{text} in one window) and \emph{sub-pattern} (i.e., the portion of the \revIII{query pattern} in one window), GenASM-DC searches for the sub-pattern within the sub-text and generates the bitvectors (\circlednumber{4}). Each processing element (PE) of GenASM-DC writes the generated bitvectors to its own dedicated SRAM, which we call TB-SRAM (\circlednumber{5}). Once GenASM-DC completes its search for the current window, GenASM-TB starts reading the stored bitvectors from TB-SRAMs (\circlednumber{6}) and generates the window's traceback output (\circlednumber{7}).} Once GenASM-TB generates this output, \rev{GenASM} computes the next window and repeats \sgii{Steps \circlednumber{3}--\circlednumber{7}} until all windows are completed.

Our hardware accelerators are designed to maximize parallelism and minimize memory footprint. Our modified GenASM-DC algorithm is highly parallelizable, and performs only simple and regular bitwise operations, so we implement \revonur{the GenASM-DC accelerator} as a systolic array based accelerator. GenASM-TB \revonur{accelerator} requires simple logic operations to perform the TB-SRAM accesses and the required control flow to complete the traceback operation. Both of our hardware accelerators are highly efficient in terms of area and power. 
We discuss \revII{them} in detail in Section~\ref{sec:bitmac-hw}.



\vspace{-5pt}
\section{G\lowercase{en}ASM-DC Algorithm} \label{sec:bitap-search}
\vspace{-3pt}

\rev{
We modify the baseline Bitap algorithm (Section~\ref{sec:background-bitap}) to (1)~enable efficient alignment of \revII{long reads}, (2)~remove the data dependency between the iterations, and (3)~provide parallelism for the large \revII{number} of iterations. 

\textbf{Long Read Support.}
\sgrev{The GenASM-DC algorithm \revII{overcomes} the word-length limit of Bitap \revIII{(Section~\ref{sec:motivation:limitations})} by} storing the bitvectors in multiple words when the query is longer than the word size. Although this modification leads to additional computation when performing shifts, it helps GenASM to support both short and long reads.
\sgrev{When shifting word~$i$ of a multi-word bitvector, the bit shifted out \revIII{(MSB)} of word~$i-1$ needs to be stored separately before performing the shift on word $i-1$. Then, that saved bit needs to be loaded as the least significant bit (LSB) of word~$i$ when the shift occurs.}  
\revIII{This causes the complexity of the algorithm to be $\lceil \frac{m}{w} \rceil \times n \times k$, where $m$ is the query length, $w$ is the word size, $n$ is the text length, and $k$ is the edit distance.}

\textbf{Loop Dependency Removal.}
In order to solve \revonur{the two-level data dependency limitation of the baseline Bitap algorithm (Section~\ref{sec:motivation:limitations}), GenASM-DC performs loop unrolling and enables computing non-neighbor (i.e., independent) bitvectors in parallel. Figure~\ref{fig:data_depend} shows an example for unrolling with four threads for text characters T0--T3 and status bitvectors R0--R7.}
\revII{\sg{For the iteration where $R[d]$ represents T2--R2} 
(i.e., \sgii{the} target cell shaded \sgii{in} dark red), $R[d-1]$ refers to T2--R1, $oldR[d-1]$ refers to T1--R1, and $oldR[d]$ refers to T1--R2 (i.e., \revIII{cells T2--R2 is dependent on}, shaded \revIII{in} light red). Based on this example, T2--R2 depends on T1--R2, T2--R1, and T1--R1, but it does not depend on T3--R1, T1--R3, or T0--R4. Thus, these independent bitvectors can be computed in parallel without waiting for \sgii{one another}.}

\begin{figure}[h!]
\centering
\vspace{-1pt}
\includegraphics[width=6.8cm,keepaspectratio]{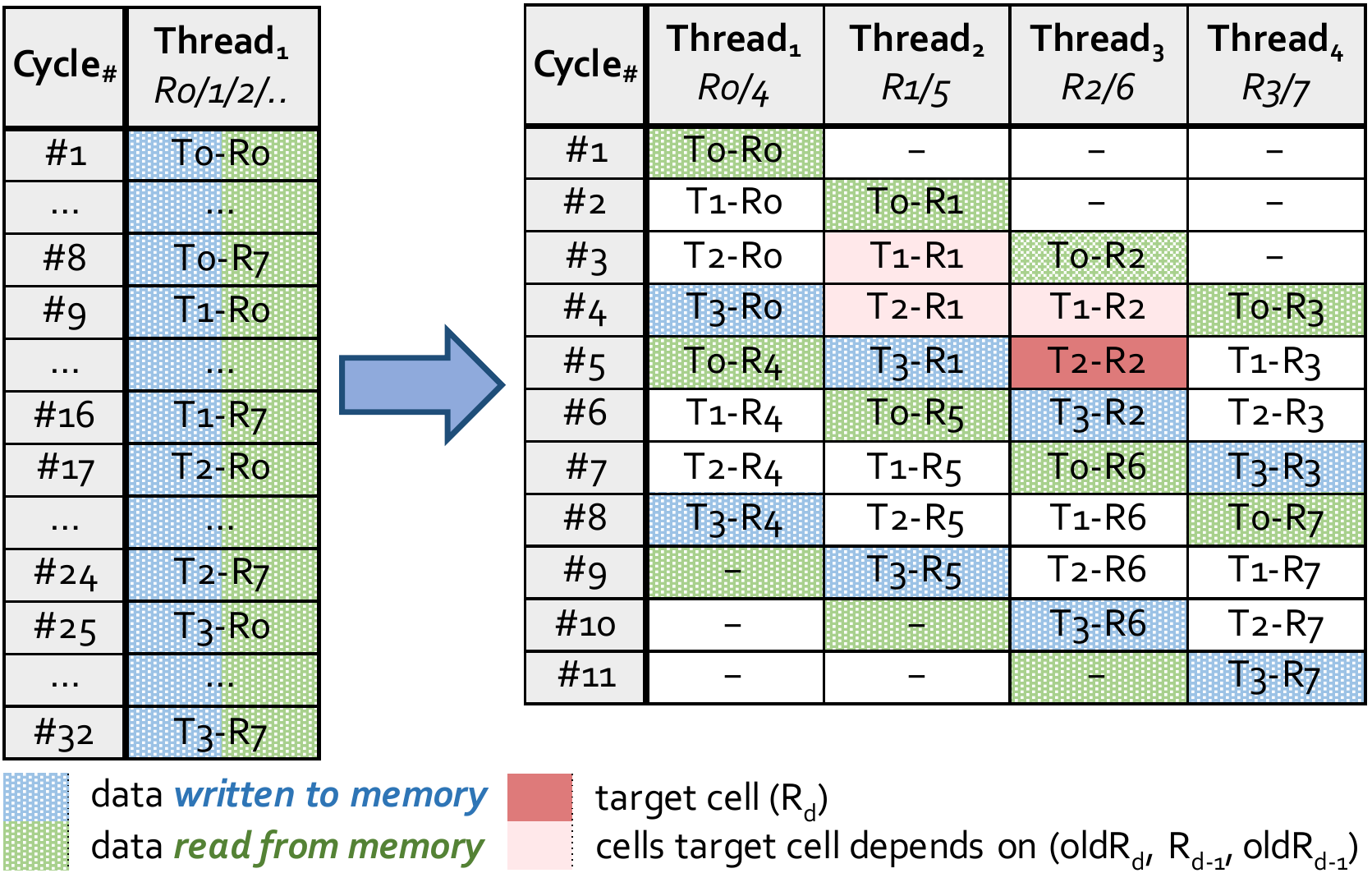}
\vspace{-4pt}
\caption{Loop unrolling in GenASM-DC.}
\label{fig:data_depend}
\vspace{-2.5pt}
\end{figure}

\textbf{Text-Level Parallelism.}
\sgrev{In addition to the parallelism enabled \revII{by} removing the loop dependencies, we enable GenASM-DC algorithm to exploit text-level parallelism. This parallelism is enabled by} dividing the text into overlapping sub-texts and searching the \revIII{query} in each of these sub-texts in parallel. 
The overlap 
ensures that we do not miss any possible match that may fall around the edges of a sub-text. To guarantee this, the overlap \sgrev{needs to} be of length $m+k$, where $m$ is the \revIII{query} length and $k$ is the edit distance threshold. 

}

\vspace{-5pt}
\section{G\lowercase{en}ASM-TB Algorithm} \label{sec:bitap-traceback}
\vspace{-3pt}

After finding the matching location of the text and the edit distance with GenASM-DC, our new traceback\revIII{~\cite{myers1988optimal,gotoh1986alignment,gotoh1982improved,miller1988sequence,waterman1984efficient,altschul1986optimal,fickett1984fast,smith1981identification,waterman1976some,ukkonen1985algorithms}} algorithm, GenASM-TB, finds the sequence of matches, substitutions, insertions and deletions, along with their positions \revV{(i.e., CIGAR string)} for the matched \revIV{region (i.e., the text region that starts from the location reported by GenASM-DC and has a length of $m+k$)}, \revIII{and reports the optimal alignment}. \revIII{Traceback execution (1)~starts from the first character of the matched region between the reference text and query pattern, (2)~examines \revIV{each character and decides} which of the four operations should be picked \revIV{in} each iteration, and (3)~ends when we reach the last character of the matched region.}
GenASM-TB uses the \revonur{intermediate bitvectors generated \revII{and saved in} each iteration of the GenASM-DC algorithm (i.e., match, substitution, deletion and insertion bitvectors generated in \revIII{Lines~15--18} in Algorithm~\ref{bitap-search-alg})}. After a \sgii{value 0} is found at the MSB of one of the $R[d]$ bitvectors \revII{(i.e., \sgii{a} \revIII{string} match is found with $d$ errors)}, 
GenASM-TB \sg{walks through the bitvectors back to the LSB, following a chain of 0s (which indicate matches at each location) and} 
reverting the bitwise operations. \revonur{\sgii{At each position, based} on which of the \revII{four} bitvectors \sgii{holds a value 0} \revII{in} each iteration (starting with \sgii{an MSB with a 0 and ending with an LSB with a 0}), the sequence of matches, substitutions, insertions and deletions (i.e., traceback output) is found for each position of the \revII{corresponding alignment found by GenASM-DC.}} Unlike GenASM-DC, GenASM-TB has an irregular \revonur{control flow} within the stored intermediate bitvectors, which depends on the \revIV{text and \revV{the} pattern.}

\begin{figure}[t!]
\centering
\vspace{-2pt}
\includegraphics[width=\columnwidth,keepaspectratio]{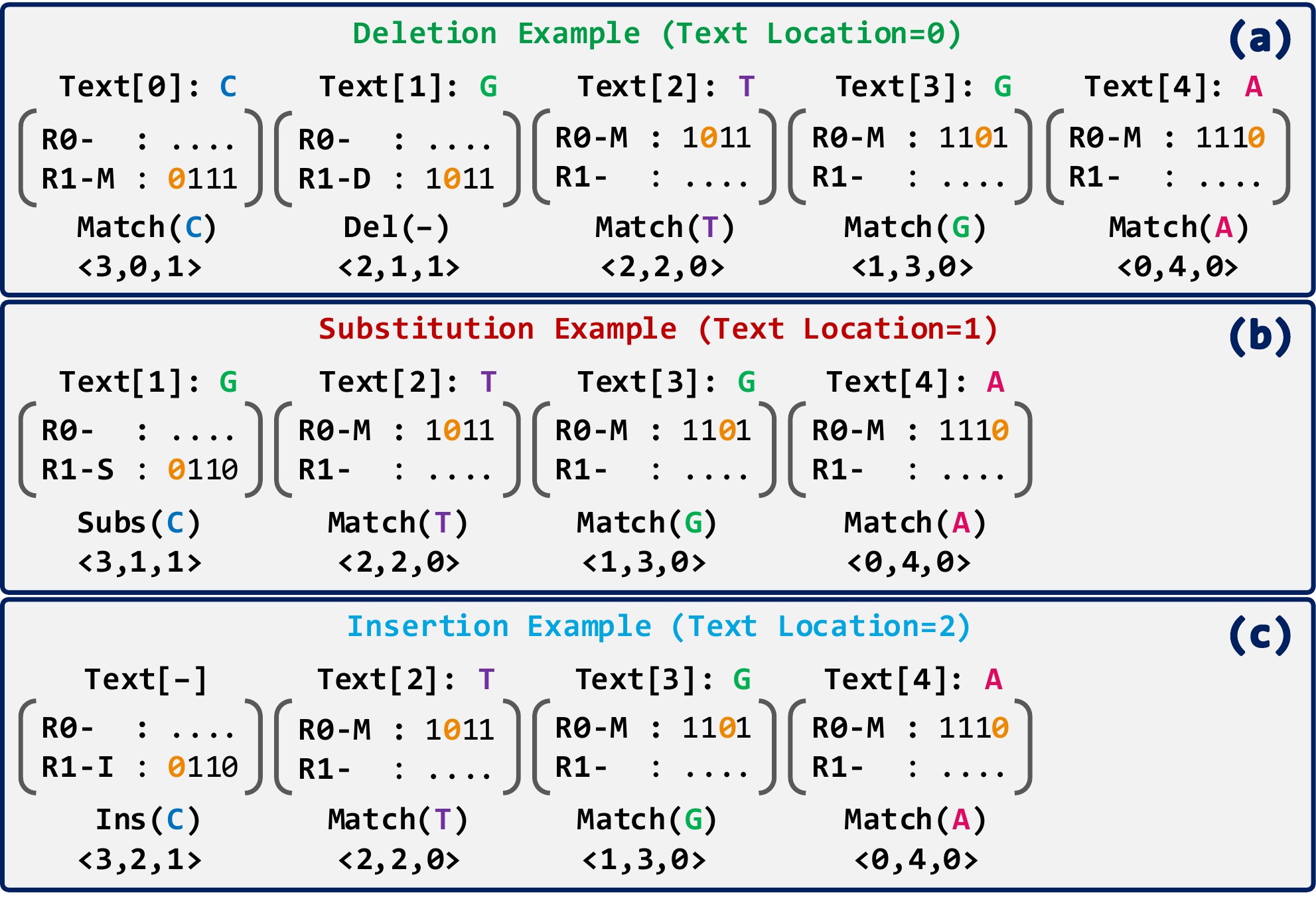}
\vspace{-14pt}
\caption{Traceback \revII{example} with GenASM-TB algorithm.} \label{fig:bitap-traceback-alg}
\vspace{-5pt}
\end{figure}

\revonur{Algorithm~\ref{bitap-traceback-alg} shows the \textit{GenASM-TB} algorithm and Figure~\ref{fig:bitap-traceback-alg} shows an example for the execution of the algorithm \revIII{for each of the alignments found in \circlednumberr{3}--\circlednumberr{5} of Figure~\ref{fig:bitap-dc-example}}.}
\revonur{In Figure~\ref{fig:bitap-traceback-alg}, <$x,y,z$> stands for \texttt{patternI}, \texttt{textI} and \texttt{curError}, respectively \revIII{(Lines~6--8 in Algorithm~\ref{bitap-traceback-alg})}. \texttt{patternI} represents the position of \sgii{a 0 currently being processed} within a given bitvector (i.e., pattern index), \texttt{textI} represents the outer loop iteration index (i.e., text index; $i$ in Algorithm~\ref{bitap-search-alg}), and \texttt{curError} represents the inner loop iteration index (i.e., number of remaining errors; $d$ in Algorithm~\ref{bitap-search-alg}). 

When we find a $0$ at \revIII{\texttt{\small{match[textI][curError][patternI]}}} (i.e., a \emph{match \revV{(M)}} is found for the current position; \revIII{Line~17}), one character \revIII{each} from both text and query is consumed, but the number of remaining errors \revII{stays the} same. Thus, the pointer moves to the next text character (as the text character \revV{is consumed}), and \sgii{the 0 currently being processed} \revII{(highlighted with orange color in Figure~\ref{fig:bitap-traceback-alg})} is right-shifted by one (as the query character is \revV{also} consumed). In other words, \texttt{textI} is incremented \revIII{(Line~28)}, \texttt{patternI} is decremented \revIII{(Line~30)}, but \texttt{curError} remains \revV{the} same. Thus, <$x,y,z$> becomes <$x-1,y+1,z$> after we find a match. For example, in Figure~\ref{fig:bitap-traceback-alg}\revII{a, for Text[0]}, we have <$3,0,1$> for the indices, and after \sgii{the match} is found, at the next position (Text[1]), we have <$2,1,1$>.

When we find a $0$ at \revIII{\texttt{\small{subs[textI][curError][patternI]}}} (i.e., a \emph{substitution \revV{(S)}} is found for the current position; \revIII{Line~19)}, one character \revII{each} from both text and query is consumed, and the number of remaining errors is decremented \revIII{(Line~26}). Thus, <$x,y,z$> becomes <$x-1,y+1,z-1$> after we find a substitution \revII{(e.g., \sgii{Text[\revIII{1}] in} Figure~\ref{fig:bitap-traceback-alg}b)}.

When we find a $0$ at \revIII{\texttt{\small{ins[textI][curError][patternI]}}} (i.e., an \emph{insertion \revV{(I)}} is found for the current position; \revV{Lines~13 and 21)}, the inserted character does not appear \revV{in the 
text}, and only a character from the pattern is consumed. \sgii{The 0 currently being processed} is right-shifted by one, but the text pointer remains the same, and the number of remaining errors is decremented. Thus, <$x,y,z$> becomes <$x-1,y,z-1$> after we find an insertion \revII{(e.g., \sgii{Text[\revIII{--}] in} Figure~\ref{fig:bitap-traceback-alg}c)}.
    
When we find a $0$ at \revIII{\texttt{\small{del[textI][curError][patternI]}}} (i.e., a \emph{deletion \revV{(D)}} is found for the current position; \revV{Lines~15 and 23)}, the deleted character does not appear \revV{in the 
pattern}, and only a character from the text is consumed. \sgii{The 0 currently being processed} is not right-shifted, but the pointer moves to the next text character, and the number of remaining errors is also decremented. Thus, <$x,y,z$> becomes <$x,y+1,z-1$> after we find an insertion \revII{(e.g., \sgii{Text[1] in} Figure~\ref{fig:bitap-traceback-alg}a)}.

}

{
\begin{algorithm}[t!]
\fontsize{7}{7}\selectfont{\revIII{\textbf{Inputs:} \texttt{text} (reference), \texttt{n}, \texttt{pattern} (query), \texttt{m}, \texttt{W} (window size), \texttt{O} (overlap size)\\
\textbf{Output:} \texttt{CIGAR} (complete traceback output)}}
\vspace{-1pt}
\caption{GenASM-TB Algorithm}\label{bitap-traceback-alg}
\begin{algorithmic}[1]
\fontsize{7}{7}\selectfont
    \State \revIII{$\texttt{<curPattern,curText>} \gets$ \texttt{<0,0>}}
    \Comment{\comm{start positions of sub-pattern and sub-text}}
    \While{\revIII{\texttt{(curPattern < m) \& (curText < n)}}}
        
        \State \revIII{$\texttt{sub-pattern} \gets \texttt{pattern[curPattern:(curPattern+W)]}$}
        \State \revIII{$\texttt{sub-text} \gets \texttt{text[curText:(curText+W)]}$}
        
        \State \revIII{$\texttt{intermediate bitvectors} \gets \texttt{GenASM-DC(sub-pattern,sub-text,W)}$}

        \State $\texttt{patternI} \gets \texttt{W-1}$ 
        \Comment{\comm{pattern index (position of 0 being processed)}}
        \State $\texttt{textI} \gets \texttt{0}$
        \Comment{\comm{text index}}
        \State $\texttt{curError} \gets \texttt{editDist from GenASM-DC}$
        \Comment{\comm{number of remaining errors}}
        \State $\revIII{\texttt{<patternConsumed,textConsumed>} \gets \texttt{<0,0>}}$
    
        \State $\texttt{prev} \gets \texttt{""}$
        \Comment{\comm{output of previous TB iteration}}
        \While{\texttt{textConsumed<(W-O) \& patternConsumed<(W-O)}}
            \State $\texttt{status} \gets \texttt{0}$
            \If{\texttt{ins[textI][curError][patternI]=0 \& prev='I'}}
                \State $\texttt{status} \gets \texttt{3; add "I" to CIGAR;}$
                \Comment{\comm{insertion-extend}}
            \ElsIf{\texttt{del[textI][curError][patternI]=0 \& prev='D'}}
                \State $\texttt{status} \gets \texttt{4; add "D" to CIGAR;}$
                \Comment{\comm{deletion-extend}}
            \ElsIf {\texttt{match[textI][curError][patternI]=0}}
                \State $\texttt{status} \gets \texttt{1; add "M" to CIGAR; } \texttt{prev} \gets \texttt{"M"}$
                \Comment{\comm{match}}
            \ElsIf{\texttt{subs[textI][curError][patternI]=0}}
                \State $\texttt{status} \gets \texttt{2; add "S" to CIGAR; } \texttt{prev} \gets \texttt{"S"}$
                \Comment{\comm{substitution}}
            \ElsIf{\texttt{ins[textI][curError][patternI]=0}}
                \State $\texttt{status} \gets \texttt{3; add "I" to CIGAR; }
                \texttt{prev} \gets \texttt{"I"}$
                \Comment{\comm{insertion-open}}
            \ElsIf {\texttt{del[textI][curError][patternI]=0}}
                \State $\texttt{status} \gets \texttt{4; add "D" to CIGAR; }
                \texttt{prev} \gets \texttt{"D"}$
                \Comment{\comm{deletion-open}}
            \EndIf
            \If {\texttt{(status > 1)}}         
                \State $\texttt{curError-{}-}$ 
                \Comment{\comm{S, D, or I}}
            \EndIf
            \If {\texttt{(status > 0) \&\& (status != 3)}}  
                \State $\texttt{textI++; textConsumed++}$
                \Comment{\comm{M, S, or D}}
            \EndIf
            \If {\texttt{(status > 0) \&\& (status != 4)}} 
                \State $\texttt{patternI-{}-; patternConsumed++}$
                \Comment{\comm{M, S, or I}}
            \EndIf
        \EndWhile
        \State $\revIII{\texttt{curPattern} \gets \texttt{curPattern+patternConsumed}}$
        \State $\revIII{\texttt{curText} \gets \texttt{curText+textConsumed}}$
    \EndWhile
\end{algorithmic}
\end{algorithm}
}

\vspace{1pt}
\textbf{\revII{Divide-and-Conquer Approach.}} Since GenASM-DC stores all of the intermediate bitvectors, 
in the worst case, the length of the text region that the query pattern maps \revonur{to} can be $m+k$, assuming all of the errors are deletions from the pattern. Since we need to store all of the bitvectors for $m+k$ characters, and compute \revV{$4 \times k$} many bitvectors within each text iteration (each $m$ bits long), for long reads with high error rates, the memory requirement becomes \textasciitilde \revV{80GB}, \revonur{when m is 10,000 and k is 1,500}. 

In order to decrease the memory footprint of the algorithm, we follow two key ideas. First, we apply a divide-and-conquer approach \rev{(similar to the tiling approach of Darwin's alignment accelerator, GACT~\cite{turakhia2018darwin})}. Instead of storing all of the bitvectors for $m+k$ text characters, we divide the text and \revIII{pattern} into overlapping windows \revonur{(i.e., sub-text and \revIII{sub-pattern; Lines~3--4 in Algorithm~\ref{bitap-traceback-alg})}} and perform the traceback computation \revII{for each window}. After all of the windows' partial traceback outputs are generated, we merge them to find the complete traceback \revV{output}. This approach helps us to decrease the memory footprint \revonur{from \revV{$((m+k) \times 4 \times k \times m)$} \revIII{bits} to \revV{$(W \times 4 \times W \times W)$ \revIII{bits}}}, where $W$ is the window size. This divide-and-conquer approach also helps us to reduce the complexity of the bitvector generation step (Section~\ref{sec:bitap-search}) from \revIII{$\lceil \frac{m}{w} \rceil \times n \times k$ to $\lceil \frac{W}{w} \rceil \times W \times W$}. 
Second, instead of storing all 4 bitvectors (i.e., match, substitution, insertion, deletion) separately, we only need to store bitvectors for match, insertion, and deletion, as the substitution bitvector can be obtained easily by left-shifting the deletion bitvector by 1 \revII{(\revIII{Line~16} in Algorithm~\ref{bitap-search-alg})}.
This modification helps us to decrease the required write bandwidth and the memory footprint 
to \revV{$(W \times 3 \times W \times W)$} \revIII{bits}.

GenASM-TB restricts the number of consumed characters from the text or the pattern to \texttt{W-O} \revonur{(\revIII{Line~11} in Algorithm~\ref{bitap-traceback-alg})} to ensure that consecutive windows share $O$ characters \revII{(i.e., overlap size)}, and thus, the traceback output can be generated accurately. 
\revIII{The sub-text and the sub-pattern corresponding to each window are found using the number of consumed text characters (\texttt{textConsumed}) and the number of consumed pattern characters (\texttt{patternConsumed}) in the previous window (Lines~31--32 in Algorithm~\ref{bitap-traceback-alg})}.


\textbf{Partial Support for Complex Scoring Schemes.} 
We extend the GenASM-TB algorithm to provide \revonur{partial} support \revIII{(Section~\ref{sec:results-accuracy})} for
non-unit costs for different edits and \revonur{the} affine gap penalty model\revonur{~\cite{gotoh1982improved,miller1988sequence,waterman1984efficient,altschul1986optimal}}. 
\revIII{By changing the order in which different traceback cases are checked in \revIII{Lines~13--24} in Algorithm~\ref{bitap-traceback-alg}, we can support different types of scoring schemes. For example, in order} \sgii{to mimic \revIII{the behavior of the} affine gap penalty \revIII{model}}, we check whether the \revonur{traceback output} that has been chosen for the previous position \revII{(i.e., \texttt{prev}) is an insertion or a deletion}. If the previous edit is a gap (insertion or deletion), and 
there is a $0$ at the current position of the insertion or deletion bitvector \revIII{(Lines~13 and 15 in Algorithm~\ref{bitap-traceback-alg})}, then we prioritize extending this previously opened gap, and choose \revII{insertion-extend or deletion-extend} as the current position's \revonur{traceback output}, depending on the type of the previous gap. \revIV{As another example}, in order to mimic the behavior of non-unit costs for different edits, we can simply sort three error cases \revII{(substitution, insertion-open, deletion-open)} from the lowest penalty to the highest penalty. \revonur{If substitutions have a lower penalty than gap openings, the order shown in Algorithm~\ref{bitap-traceback-alg} should remain the same. However, if substitutions have a greater penalty than gap openings, we should check for the substitution case after checking the insertion-open and deletion-open cases \revIII{(i.e., Lines~19--20 should come after Line~24 in Algorithm~\ref{bitap-traceback-alg})}.}

\vspace{-6pt}
\section{G\lowercase{en}ASM Hardware Design} \label{sec:bitmac-hw}
\vspace{-3pt}

\begin{figure*}[b!]
\centering
\vspace{1pt}
\includegraphics[width=17.3cm,keepaspectratio]{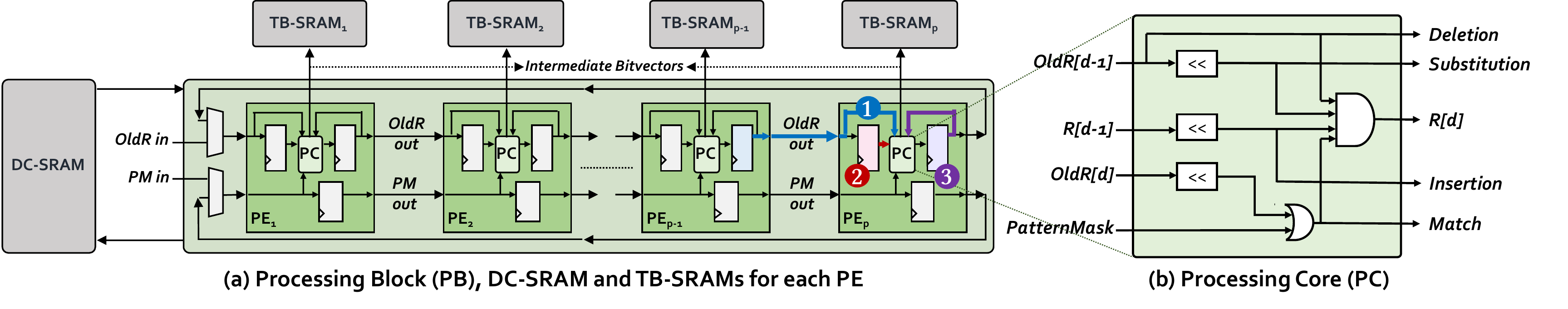}
\vspace{-5pt}
\caption{\revonur{Hardware design of GenASM-DC.}}
\label{fig:bitmac-dc-pb}
\vspace{-7pt}
\end{figure*}

\textbf{GenASM-DC Hardware.} We implement GenASM-DC as a linear cyclic systolic array\revIII{~\cite{kung1978systolic,kung1982systolic}} based accelerator.
The accelerator is optimized to reduce both the memory bandwidth and the memory footprint. Feedback logic enabling cyclic systolic behavior allows us to fix the required number of memory ports\revIII{~\cite{kung1982systolic}} and to reduce memory footprint. 


A GenASM-DC accelerator consists of a processing block (PB; Figure~\ref{fig:bitmac-dc-pb}a) along with \revonur{a} control and memory management logic.
A PB consists of multiple processing elements (PEs). Each PE contains a single processing core (PC; Figure~\ref{fig:bitmac-dc-pb}b) and flip-flop-based storage logic. The PC is the primary compute unit, and implements
\revIII{Lines 15--19 of Algorithm~\ref{bitap-search-alg}} to perform the approximate string matching for a $w$-bit query pattern.
The number of PEs in a PB is based on compute, area, memory bandwidth and power requirements. This block also implements the logic to load data from outside of the array \revIII{(i.e., DC-SRAM; Figure~\ref{fig:bitmac-dc-pb}a)} or internally for cyclic operations.

GenASM-DC uses two types of \revonur{SRAM buffers (Figure~\ref{fig:bitmac-dc-pb}a)}: (1)~DC-SRAM, \revonur{which stores the reference text, the pattern bitmasks for the query read, and the intermediate data generated from PEs (i.e., $oldR$ values and MSBs required for shifts; \revIII{Section~\ref{sec:bitap-search}});}
and (2)~TB-SRAM, \revonur{which stores} the intermediate bitvectors \revonur{from GenASM-DC for later use by} GenASM-TB. For a 64-PE configuration with 64~bits of processing per PE, \revIII{and} \revonur{for the case where we have a long (10Kbp) read\footnote{\revIII{Although we use 10Kbp-long reads in our analysis (Section~\ref{sec:methodology:datasets}), GenASM does \emph{not} have any limitation on the length of \revIV{reads as a result of} our divide-and-conquer approach (Section~\ref{sec:bitap-traceback}).}} with a high error rate (15\%) and a corresponding text region of 11.5Kbp}, GenASM-DC requires a total of \revonur{8KB DC-SRAM storage.} 
For each PE, we have a dedicated \revIII{TB-SRAM, 
which} stores the match, insertion and deletion bitvectors generated by the corresponding PE. For the same configuration of GenASM-DC, 
each PE requires a total of 1.5KB TB-SRAM storage, with a single R/W port. In each cycle, 192 bits of data (24B) is written to each TB-SRAM by each PE.

\sgii{When each thread (i.e., each column) in Figure~\ref{fig:data_depend} is mapped to a PE, GenASM-DC coordinates the data dependencies across DC iterations, with the help of two flip-flops in each PE. For example, T2--R2 in Figure~\ref{fig:data_depend} is generated by $PE_x$ in $Cycle_y$, and is mapped to $R[d]$. In order to generate T2--R2, T2--R1 (which maps to $R[d-1]$) needs to be generated by $PE_{x-1}$ in $Cycle_{y-1}$ (\circlednumber{1} in Figure~\ref{fig:bitmac-dc-pb}), T1--R1 (which maps to $oldR[d-1]$) needs to be generated by $PE_{x-1}$ in $Cycle_{y-2}$ (\circlednumber{2}), and T1--R2 (which maps to $oldR[d]$) needs to be} generated by $PE_{x}$ in $Cycle_{y-1}$ (\circlednumber{3}), where $x$ is the PE index and $y$ is the cycle index. 
With this dependency-aware mapping, regardless of the number of instantiated PEs, \revV{we can successfully limit DC-SRAM traffic for a single PB 
to only one read and one write per cycle.}


\textbf{GenASM-TB Hardware.}
After GenASM-DC finishes writing all of the \revV{intermediate} bitvectors to TB-SRAMs, GenASM-TB \revIII{reads} them by following an irregular control flow, which depends on the \revV{text and the pattern} to find the optimal alignment \revIII{(by \revIV{implementing} Algorithm~\ref{bitap-traceback-alg})}.

\begin{figure}[h!] 
\centering
\vspace{-2pt}
\includegraphics[width=\linewidth,keepaspectratio]{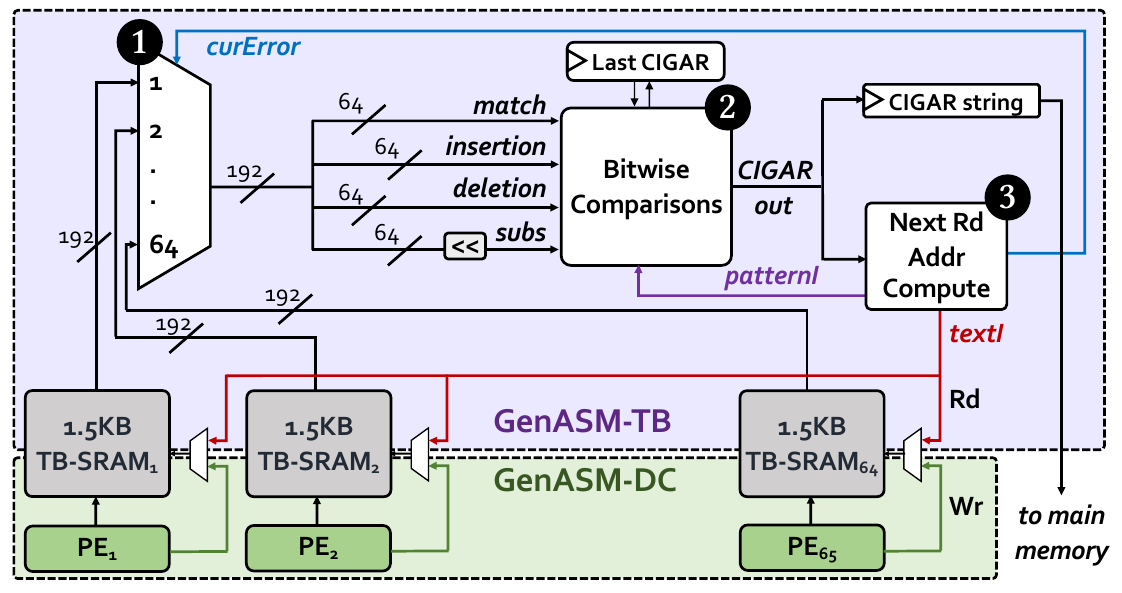}
\vspace{-15pt}
\caption{\revonur{Hardware design of GenASM-TB.
}} \label{fig:bitmac-tb-hw-overall}
\vspace{0pt}
\end{figure}

\revonur{In our GenASM configuration, where we have 64 PEs and 64~bits per PE in a GenASM-DC accelerator, and the window size ($W$) is 64 (Section~\ref{sec:bitap-traceback}), we have one 1.5KB TB-SRAM (which fits \revIII{our} 24B/cycle $\times$ 64 cycles/window \revIII{output storage requirement}) for each of the 64 PEs.}
As Figure~\ref{fig:bitmac-tb-hw-overall} shows, 
a single GenASM-TB accelerator is connected to all of these 64 TB-SRAMs (96KB, in total). 
In each GenASM-TB cycle, we read from only one TB-SRAM. \revIII{\texttt{curError} provides the index of the TB-SRAM that we read from; \texttt{textI} provides the starting index within this TB-SRAM, which we read the next set of bitvectors from; and \texttt{patternI} provides the position of the 0 being processed (Algorithm~\ref{bitap-traceback-alg}).}

We implement the GenASM-TB hardware using very simple logic (Figure~\ref{fig:bitmac-tb-hw-overall}), which \revIII{\circlednumber{1}~reads} the bitvectors from one of the TB-SRAMs using the computed address, \revIII{\circlednumber{2}~performs} the required \revIII{bitwise comparisons} to find the CIGAR character for the current position, and \revIII{\circlednumber{3}~computes} the next TB-SRAM address to read the new set of bitvectors. After GenASM-TB finds the complete CIGAR string, it writes the output to main memory and completes its \revIII{ execution. 
}

\textbf{Overall System.}
\label{sec:overall-system}
\revIII{We design our system to take advantage of 
modern 3D-stacked memory systems~\cite{ghose2019demystifying, kim2016ramulator}, 
such as the Hybrid Memory Cube (HMC)~\cite{hmc}
or High-Bandwidth Memory (HBM)~\cite{hbm, lee2016simultaneous}.
Such memories are made up of multiple layers of DRAM arrays that
are stacked vertically in a single package.
These layers} are connected via high-bandwidth links called
\emph{through-silicon vias} (TSVs) 
\revIII{that provide lower-latency and more energy-efficient data access to the layers than the external DRAM I/O pins~\cite{davis2005demystifying,lee2016simultaneous}.
Memories such as HMC and HBM include a dedicated \emph{logic layer}
that connects to the TSVs and allows processing elements to
be implemented in memory to exploit the efficient data access.
Due to thermal and area constraints, only simple processing elements that execute low-complexity operations (e.g., bitwise logic, simple arithmetic, simple cores) can be included in the logic layer~\cite{boroumand2018google,drumond2017mondrian,tetris,ahn2015pim,ahn2015scalable,hsieh2016accelerating,hsieh2016transparent,mutlu2019processing,boroumand2019conda,pattnaik2016scheduling,Kim2018}.}

\revIII{We decide to implement GenASM in the logic layer of 3D-stacked memory, for two reasons.}
First, we can exploit the natural subdivision within 3D-stacked memory (e.g., vaults in HMC\revIII{~\cite{hmc}, pseudo-channels in HBM~\cite{hbm}}) to efficiently enable parallelism across multiple GenASM accelerators. This subdivision allows accelerators to work in parallel without interfering with each other. Second, we can reduce the power consumed for DRAM accesses by reducing off-chip data movement across the memory channel\revIII{~\cite{mutlu2019processing}}.
\revIII{Both} of our hardware accelerators are highly efficient in terms of area and power (Section~\ref{sec:results:area-power}), and can fit within the 
\revIII{logic layer's constraints}.

\revIII{To illustrate how GenASM takes advantage of 3D-stacked memory, we discuss an example implementation of GenASM inside the logic layer of a 16GB HMC with 32~vaults~\cite{hmc}.}
Within each vault, the logic layer contains a GenASM-DC accelerator, its associated DC-SRAM (8KB), a GenASM-TB accelerator, and TB-SRAMs (64$\times$1.5KB). Since we have small SRAM buffers for both DC and TB to exploit locality, GenASM accesses the memory and utilizes the memory bandwidth only to read the reference and the query sequences. \revmicro{One GenASM accelerator at each vault requires 105--142 MB/s bandwidth, thus the total bandwidth requirement of all 32 GenASM accelerators is 3.3--4.4 GB/s \revIII{(which is much less than peak bandwidth provided by modern 3D-stacked memories)}. 
}
\section{G\lowercase{en}ASM Framework} 
\label{sec:bitmac-framework}
\vspace{-2pt}

\rev{
We demonstrate the efficiency and flexibility of the GenASM acceleration framework by describing three use cases of approximate string matching in genome sequence analysis: (1)~read alignment step of short and long read mapping, (2)~pre-alignment filtering for short reads, and (3)~edit distance calculation between \revIII{any two 
sequences}. We believe \revonurcan{the} GenASM framework can be useful for many other use cases, and we discuss some of them briefly in Section~\ref{sec:bitmac-framework-other}.

\textbf{Read Alignment of Short and Long Reads.}\label{sec:bitmac-framework-aln}
As we explain in Section~\ref{sec:background:pipeline}, read alignment is the last step of short and long read mapping. In read alignment, all of the \revV{remaining} \revonur{candidate mapping \revIII{regions of the reference genome} and the \revIII{query reads} are aligned,} in order to identify the mapping that yields \revonur{either the lowest total number of errors (if using edit distance based scoring) or the highest score (if using \revV{a user-defined scoring function}).}
Thus, read alignment can be a use case for approximate string matching, \revonurcan{since} errors (i.e., substitutions, insertions, deletions) should be \revonur{taken into account} when aligning the sequences. As part of read alignment, we also need to generate the traceback output for the best alignment between \revonur{the reference \revIII{region and the read.}}

For read alignment, \revonurcan{the} whole GenASM pipeline, \revonurcan{as} explained in Section~\ref{sec:bitmac-overall}, should be executed, including the traceback step. In general, read alignment requires more complex scoring schemes, where different types of edits have non-unit costs. Thus, GenASM-TB should be configured based on the given cost \revonurcan{of} each type of edit (Section~\ref{sec:bitap-traceback}). As GenASM framework can work with arbitrary length sequences, we can use it to accelerate both short read and long read alignment.

\textbf{Pre-Alignment Filtering for Short Reads.}\label{sec:bitmac-framework-filter}
In the pre-alignment filtering step of short read mapping, the candidate \revII{mapping} locations, reported by the seeding step, are further filtered by using different mechanisms. 
\sgii{Although the regions of the reference at these candidate mapping locations share common seeds with query reads,}
they are not necessarily \emph{similar} sequences. 
To avoid examining dissimilar sequences at the \revonurcan{downstream} computationally-expensive read alignment step, \revonurcan{a} pre-alignment filter estimates the edit distance between \revonur{every read} and \revIII{the regions of the reference at each read's candidate mapping locations}, and \revonurcan{uses} this estimation to quickly decide whether or not read alignment is needed. If the sequences are dissimilar enough, significant amount of time is saved by avoiding the expensive alignment step\revonurcan{~\cite{gatekeeper, alser2019sneakysnake, Alser2019, Xin2013, Xin2015}}.

\revonur{
In pre-alignment filtering, since we only need to estimate \revIII{(approximately)} the edit distance and check whether it is above a user-defined threshold, GenASM-DC can be used as a pre-alignment filter. As GenASM-DC is very efficient when we have shorter sequences and a low error threshold (due to \revonur{the} $O(m \times n \times k)$ complexity of the underlying Bitap algorithm, where $m$ is the query length, $n$ is the reference length, and $k$ is the number of allowed errors), GenASM framework can \revonur{efficiently} accelerate the pre-alignment filtering step of \revIII{especially} short read mapping.\footnote{\revIII{Although we believe that GenASM can also be used as a pre-alignment filter for long reads, we leave the evaluation of this use case for future work.}}
}

\vspace{1pt}
\textbf{Edit Distance Calculation.}\label{sec:bitmac-framework-edc}
Edit distance, also called Levenshtein distance~\cite{levenshtein1966binary}, is the minimum number of edits (i.e., substitutions, insertions and deletions) required to convert one sequence to another. Edit distance calculation is one of the fundamental operations in genomics to measure the similarity or distance between two sequences~\cite{vsovsic2017edlib}. As we explain in Section~\ref{sec:background-bitap}, the Bitap algorithm, which is the underlying algorithm of GenASM-DC, is \revonur{originally} designed for edit distance calculation. Thus, GenASM framework can accelerate edit distance calculation between any \revonurcan{two} arbitrary-length genomic sequences.

Although GenASM-DC can find the edit distance by itself and \revIII{traceback 
is} optional for this use case, DC-TB interaction is required \revonurcan{in our accelerator} \revonur{to exploit} the efficient divide-and-conquer approach GenASM follows. \revonur{Thus,} 
GenASM-DC and GenASM-TB work together to find the minimum edit distance in a fast and memory-efficient way, but the traceback output \revonur{is not generated or reported by default (though it can optionally be enabled)}.

}
\vspace{-6pt}
\section{Evaluation Methodology} \label{sec:methodology}
\vspace{-3pt}

\textbf{Area and Power Analysis.}
We synthesize \revonur{and place \& route} the GenASM-DC and GenASM-TB accelerator datapaths using \revonur{the} Synopsys Design Compiler~\cite{synopsysdc} with a typical 28nm \revonurcan{low-power} process, with memories generated using an \revonurcan{industry-grade} SRAM compiler, to analyze the accelerators' area and power.
Our synthesis targets \revIV{post-routing} timing closure at \hl{1GHz} clock frequency.
We then use \revonur{an in-house cycle-accurate} simulator parameterized with the synthesis and memory estimations to drive the performance and \revonur{power} analysis.



We evaluate a 16GB HMC-like 3D-stacked DRAM architecture, with 32 vaults~\cite{hmc} and 256GB/s of internal bandwidth~\cite{boroumand2018google,hmc},
and a clock frequency of 1.25GHz~\cite{hmc}. 
The amount of available area in the logic layer for GenASM is around 3.5--4.4 mm\textsuperscript{2} per vault~\cite{drumond2017mondrian,boroumand2018google}. The power budget of our PIM logic per vault is 312mW~\cite{drumond2017mondrian}.

\textbf{Performance Model.}
We build a \revonur{spreadsheet-based} analytical model \revonur{for GenASM-DC and GenASM-TB, which} considers reference \revV{genome} (i.e., text) length, query \revonur{read} (i.e., pattern) length, maximum edit distance, window size, \revV{hardware design} parameters (number of PEs, bit width of \revonurcan{each} PE) and number of vaults as input parameters and projects compute cycles, DRAM \revonur{read/write bandwidth, SRAM read/write} bandwidth, and memory footprint. \revonur{We verify the analytically-estimated} cycle counts for various PE configurations with the cycle counts collected from \revonurcan{our} RTL simulations.  


\revonur{\textbf{Read Alignment Comparisons.}}
For the read alignment \revonur{use case}, we compare GenASM with the read alignment steps of two \revonurcan{commonly-used} state-of-the-art read mappers: Minimap2~\cite{li2018minimap2} and BWA-MEM~\cite{li2013aligning}, running on \revonur{an} Intel\textsuperscript{\textregistered} Xeon\textsuperscript{\textregistered} Gold 6126 CPU\revonur{~\cite{intel_cpu}} operating at 2.60GHz, with 64GB DDR4 memory. Software baselines are run with a single thread and with 12 threads.
We measure the execution time and power consumption of 
the alignment steps in Minimap2 and BWA-MEM.  We measure the individual power consumed by each tool using Intel's PCM power utility~\cite{intelpcm}. 

\revmicro{We also compare GenASM with \revII{a} state-of-the-art GPU-accelerated short read alignment tool, GASAL2~\cite{gasal2}.  We run GASAL2 on an Nvidia Titan V GPU\revonur{~\cite{nvidia_titan}} with 12GB HBM2 memory~\cite{hbm}. To fully utilize the GPU, we configure the number of alignments per batch based on the GPU's number of multiprocessors and the maximum number of threads per multiprocessor, as described in the GASAL2 paper~\cite{gasal2}. To better analyze the high parallelism that \revonur{the GPU} provides, we replicate our datasets to obtain datasets with 100K, 1M and 10M reference-read pairs for short reads. We run the datasets with GASAL2, and collect kernel time and average power consumption using \emph{nvprof}~\cite{nvprof}.}

\par

We also compare GenASM with two state-of-the-art \revonur{hardware-based} alignment accelerators, GACT of Darwin \cite{turakhia2018darwin} and SillaX of GenAx \cite{fujiki2018genax}. 
We synthesize and execute the open-source RTL for GACT~\cite{darwingithub}.
We estimate the performance of SillaX using data reported by the original work~\cite{fujiki2018genax}.

We \sgii{analyze the alignment accuracy of GenASM} by comparing the alignment outputs \sgii{(i.e., alignment score, edit distance, and CIGAR string)} of GenASM with the alignment outputs of BWA-MEM \revII{and Minimap2, for short reads and long reads, respectively.}
We obtain the BWA-MEM and Minimap2 alignments by running the tools with their default settings.

\revonur{\textbf{Pre-Alignment Filtering Comparisons.} We} compare 
GenASM with Shouji~\cite{Alser2019}, which is \revonur{the state-of-the-art} FPGA-based pre-alignment filter for short reads. \revonur{For execution time and filtering accuracy \revonurcan{analyses}, we use data reported by the original work~\cite{Alser2019}. For power analysis, we report the total power consumption of Shouji using the power analysis tool in Xilinx Vivado~\cite{vivado}, after synthesizing and implementing the open-source FPGA design of Shouji~\cite{shoujigithub}.}

\revonur{\textbf{Edit Distance Calculation Comparisons.} We compare}
GenASM with the state-of-the-art \revonur{software-based} read alignment library, Edlib~\cite{vsovsic2017edlib}, \revonur{running on an Intel\textsuperscript{\textregistered} Xeon\textsuperscript{\textregistered} Gold 6126 CPU~\cite{intel_cpu} operating at 2.60GHz, with 64GB DDR4 memory.} Edlib uses \revonur{the} Myers' bitvector algorithm~\cite{myers1999fast} to find the edit distance between two sequences. We use the default global \revonur{Needleman-Wunsch (NW)~\cite{needleman1970general}} mode of Edlib to perform our comparisons. \revonur{We measure the power consumed by Edlib using Intel's PCM power utility~\cite{intelpcm}.}

\revonur{We also compare GenASM with ASAP~\cite{Banerjee2019}, which is the state-of-the-art FPGA-based} \revonurcan{accelerator for computing the edit distance between two short reads.} \revonur{We estimate the performance of ASAP using data reported by the original work~\cite{Banerjee2019}.}

\textbf{Datasets.}
\label{sec:methodology:datasets}
For the read alignment use case, we evaluate GenASM using the latest major release of the human genome assembly, GRCh38 \cite{ncbi38genome}. We use the 1--22, X, and Y chromosomes by filtering the unmapped contigs, unlocalized contigs, and mitochondrial genome.
Genome characters are encoded into 2-bit patterns
(A = 00, C = 01, G = 10, T = 11). 
With this encoding, the reference genome uses \SI{715}{\mega\byte} of memory.

We generate four sets of long reads \revV{(i.e., PacBio and ONT datasets)} using PBSIM~\cite{ono2012pbsim} and three sets of short reads \revV{(i.e., Illumina datasets)} using Mason~\cite{holtgrewe2010mason}.
For the PacBio datasets, we use the default error profile for the continuous long reads (CLR) in PBSIM. For the ONT datasets, we modify the settings to match the error profile of ONT reads sequenced using R9.0 chemistry~\cite{jain2017minion}. Both datasets have 240,000 reads of length 10Kbp, \revonur{each simulated with 10\% and 15\% error rates}. The Illumina datasets have \hl{200,000 reads of length 100bp, 150bp, and 250bp}, \revonur{each simulated with a 5\% error rate}.

For the pre-alignment filtering use case, we use two datasets that Shouji~\cite{Alser2019} provides as test cases: \revIII{reference-read pairs (1)~of length 100bp with an edit distance threshold of 5, and (2)~of length 250bp with an edit distance threshold of 15.}

For the edit distance calculation use case, we use the publicly-available dataset that Edlib~\cite{vsovsic2017edlib} \revV{provides. 
The} dataset includes two real DNA sequences, which are 100Kbp and 1Mbp in \revonur{length, and artificially-mutated versions of the original DNA sequences with measures of similarity ranging between 60\%--99\%. Evaluating this set of sequences with varying values of similarity and length enables us to demonstrate how these parameters affect performance.}

\vspace{-5pt}
\section{Results} \label{sec:results}

\vspace{-2pt}

\subsection{Area and Power Analysis}\label{sec:results:area-power}
\vspace{-2pt}
Table \ref{table:area} shows the area and power breakdown of each component in GenASM, \revonur{and the total area overhead and power consumption of (1)~a single GenASM accelerator (in 1 vault) and (2)~32 GenASM accelerators (in 32 vaults). Both GenASM-DC and GenASM-TB \revonur{operate} at 1GHz.}

\revonur{The area} overhead of \revonur{one} GenASM accelerator is \SI{0.334}{\milli\meter\squared}, and \revonur{the} power consumption of \revonur{one} GenASM accelerator, including the SRAM power, is \SI{101}{\milli\watt}. \revonur{When we compare GenASM with a \revonur{single core of a} modern Intel\textsuperscript{\textregistered} Xeon\textsuperscript{\textregistered} Gold 6126 CPU\revonur{~\cite{intel_cpu}} (which we conservatively estimate to use \SI{10.4}{\watt}~\cite{intel_cpu} and \SI{32.2}{\milli\meter\squared}~\cite{skylake-diearea} per core), we find that GenASM is significantly more efficient in terms of both area and power consumption.}
As we have one GenASM accelerator \revonur{per vault}, 
the total area overhead of \revonur{GenASM in all 32 vaults is} \SI{10.69}{\milli\meter\squared}. 
Similarly, the total power consumption of \revonur{32 GenASM accelerators is} \SI{3.23}{\watt}.

\begin{table}[h!]
\small
\centering
\caption{Area and power breakdown of GenASM.}
\label{table:area}
\vspace{-4.5pt}
\includegraphics[width=6.8cm,keepaspectratio]{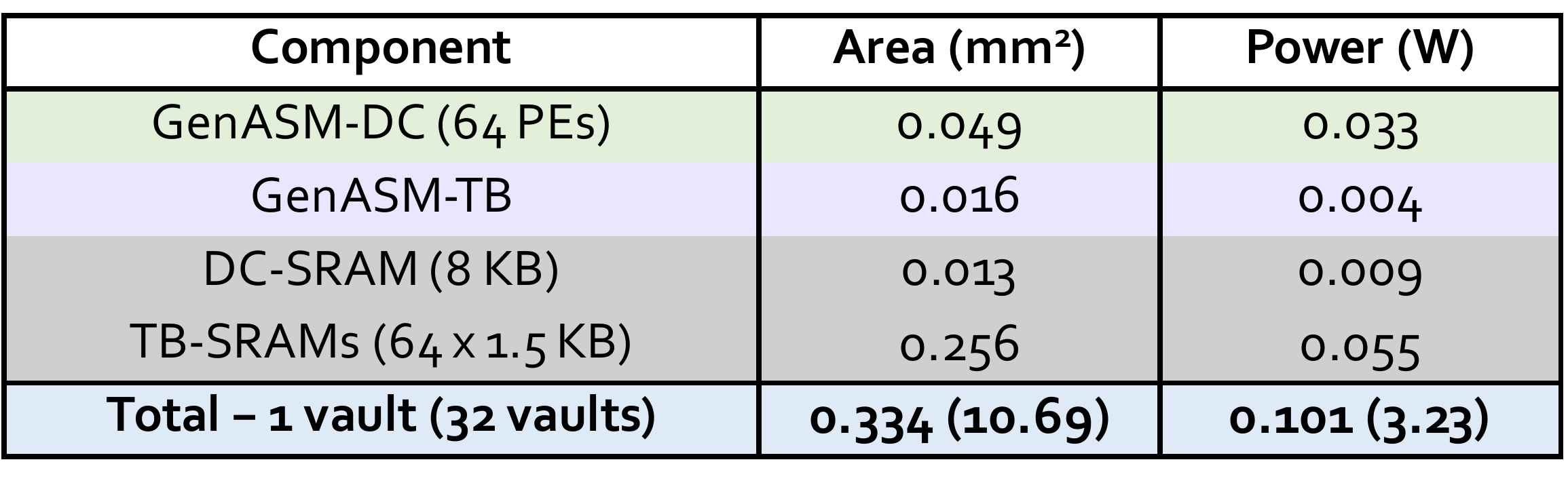}
\vspace{0pt}
\end{table}



\vspace{-2pt}
\subsection{Use Case 1: Read Alignment}\label{sec:results-alignment}
\begin{figure}[h!]
\centering
\vspace{-3pt}
\includegraphics[width=\columnwidth,keepaspectratio]{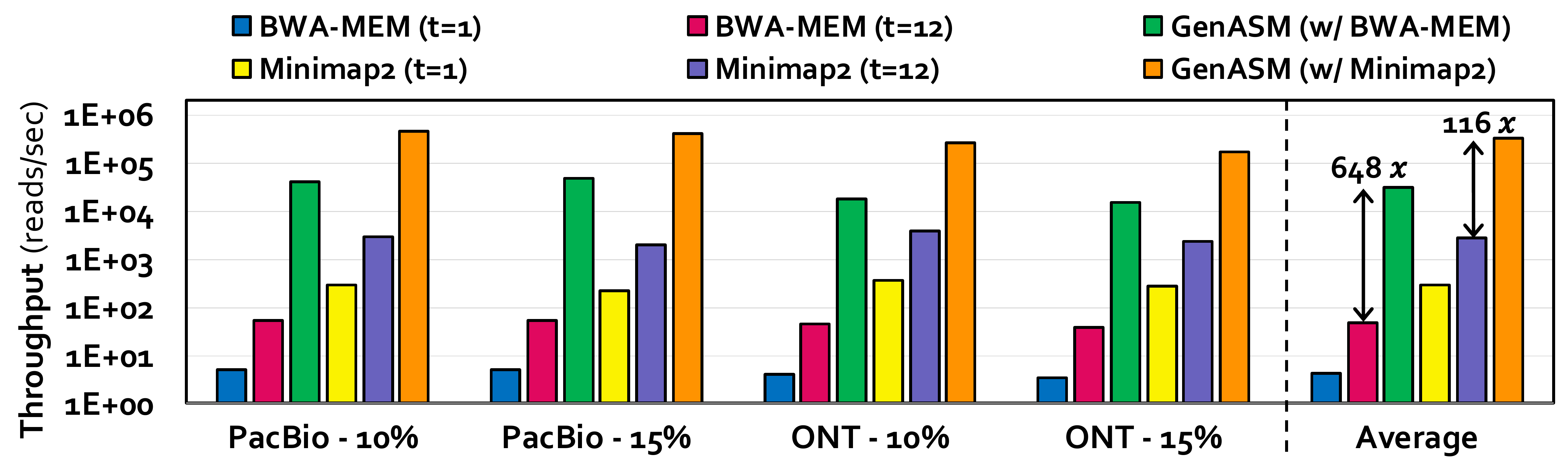}
\vspace{-15pt}
\caption{\revonur{Throughput comparison of GenASM and the alignment steps of BWA-MEM and Minimap2 for long reads.}} \label{fig:throughput-result-long}
\vspace{0pt}
\end{figure}

\vspace{-2pt}
\textbf{Software Baselines (CPU).} Figure~\ref{fig:throughput-result-long} shows the read alignment throughput (reads/sec) of GenASM and the alignment steps 
of BWA-MEM and Minimap2, when aligning long noisy PacBio and ONT reads against the human reference genome. 
When comparing with BWA-MEM, we run GenASM with the candidate locations reported by BWA-MEM's filtering step. Similarly, when comparing with Minimap2, we run GenASM with the candidate locations reported by Minimap2's filtering step. GenASM's throughput is determined by the throughput of the execution of GenASM-DC and GenASM-TB with window size ($W$) \revIII{of} 64 and overlap size ($O$) \revIII{of} 24.

As Figure~\ref{fig:throughput-result-long} shows, GenASM \revonur{provides (1)}~\hl{$7173\times$} and \hl{$648\times$} throughput improvement over the alignment step of BWA-MEM for its \revIII{single-thread} and \revIII{12-thread} execution, respectively, \revonur{and (2)~}\hl{$1126\times$} and \hl{$116\times$} throughput improvement over the alignment step of Minimap2 for its \revIII{single-thread} and \revIII{12-thread} execution, respectively. 

Based on our power analysis with long reads, we find that power consumption of BWA-MEM's alignment step is \SI{58.6}{\watt} and \SI{109.5}{\watt}, and power consumption of Minimap2's read alignment step is \SI{59.8}{\watt} and \SI{118.9}{\watt} for their \revIII{single-thread} and \revIII{12-thread} executions, respectively. \revonur{GenASM consumes only 3.23W, and thus reduces the power consumption of} the alignment steps of BWA-MEM and Minimap2 by $18\times$ and $19\times$ for single-thread execution, and by $34\times$ and $37\times$ for 12-thread execution, respectively. 


\begin{figure}[b!]
\centering
\vspace{-4.5pt}
\includegraphics[width=\columnwidth,keepaspectratio]{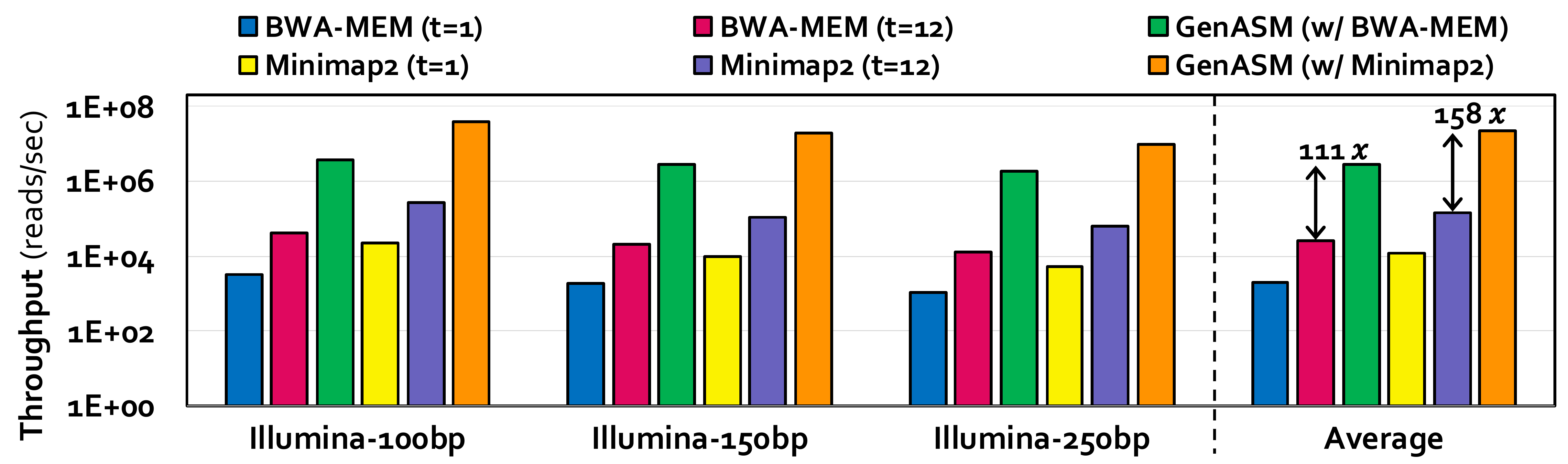}
\vspace{-15pt}
\caption{Throughput comparison of GenASM and the alignment steps of BWA-MEM and Minimap2 for short reads.} \label{fig:throughput-result-short}
\vspace{-2pt}
\end{figure}

Figure~\ref{fig:throughput-result-short} \revonur{compares} the read alignment throughput (reads/sec) of GenASM \revonur{with that of} the alignment steps \revonur{of BWA-MEM and Minimap2, when aligning short} Illumina reads against the human reference genome. 
\revonur{GenASM provides (1)~}\hl{$1390\times$} and \hl{$111\times$} throughput improvement over the alignment step of BWA-MEM for its \revIII{single-thread} and \revIII{12-thread} execution, respectively, \revonur{and (2)~}\hl{$1839\times$} and \hl{$158\times$} throughput improvement over the alignment step of Minimap2 for its \revIII{single-thread} and \revIII{12-thread} execution. 

Based on our power analysis with short reads, we find that GenASM reduces the power consumption over the alignment steps of BWA-MEM and Minimap2 by $16\times$ and $18\times$ for \revIII{single-thread execution}, and by $33\times$ and $31\times$ for \revIII{12-thread} \revonur{execution}, respectively. 

\begin{figure}[h!]
\centering
\vspace{-2pt}
\includegraphics[width=\columnwidth,keepaspectratio]{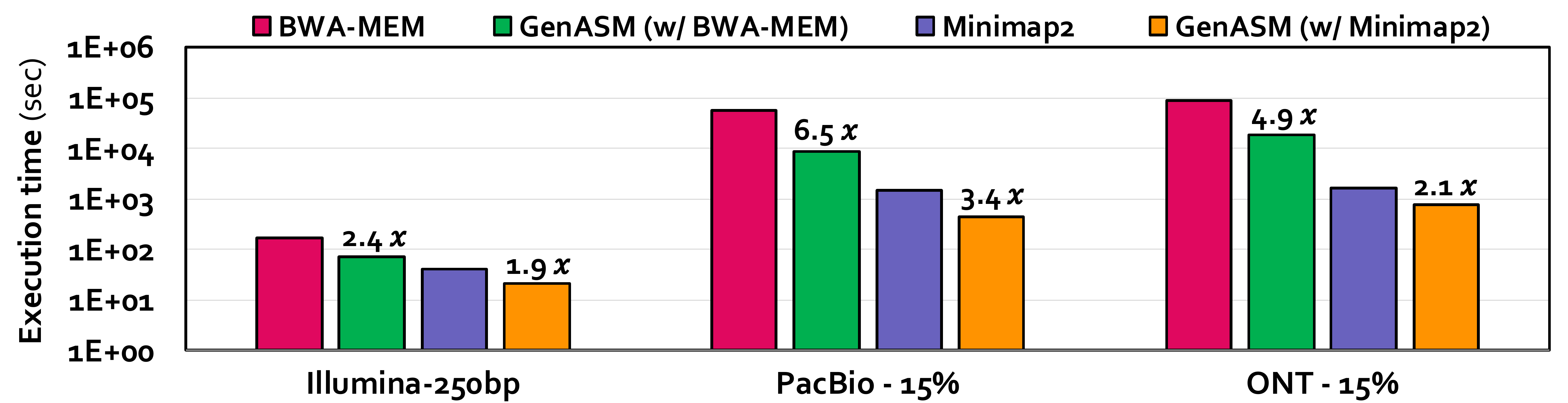}
\vspace{-15pt}
\caption{\revIII{Total execution time of \revonur{the entire} BWA-MEM and Minimap2 pipelines with and without GenASM.}} \label{fig:fullpipeline-mixed}
\vspace{0pt}
\end{figure}

Figure~\ref{fig:fullpipeline-mixed} shows the total execution time of \revonur{the entire} BWA-MEM and Minimap2 pipelines, 
\revonur{along with the total execution time when the alignment steps of each pipeline are replaced by GenASM,}
for the three representative input datasets.
\revIV{As Figure~\ref{fig:fullpipeline-mixed} shows, GenASM provides (1)~\hl{$2.4\times$} and \hl{$1.9\times$} speedup for Illumina reads (250bp); (2)~\hl{$6.5\times$} and \hl{$3.4\times$} speedup for PacBio reads (15\%); and (3)~\hl{$4.9\times$} and \hl{$2.1\times$} speedup for ONT reads (15\%), over the \revonur{entire} pipeline executions of BWA-MEM and Minimap2, respectively.}

\textbf{Software Baselines (GPU).}
We compare GenASM with \revonur{the state-of-the-art GPU aligner,} GASAL2~\cite{gasal2}, \revonur{using three datasets of varying size (100K, 1M, and 10M reference-read pairs)}. \revIV{Based on our analysis,} we \revonur{make three findings.
First, for} 100bp Illumina reads, GenASM provides \revonur{9.9$\times$, 9.2$\times$, and 8.5$\times$} speedup over GASAL2, while reducing the power consumption by \revonur{15.6$\times$, 17.3$\times$ and 17.6$\times$} for 100K, 1M, and 10M datasets, respectively.
\revonur{Second, for} 150bp Illumina reads, GenASM provides \revonur{15.8$\times$, 13.1$\times$, and 13.4$\times$} speedup over GASAL2, while reducing the power consumption by \revonur{15.4$\times$, 18.0$\times$, and 18.7$\times$} for 100K, 1M, and 10M datasets, respectively.
\revonur{Third, for} 250bp Illumina reads, GenASM provides 21.5$\times$, 20.6$\times$, and 21.1$\times$ speedup over GASAL2, while reducing the power consumption by 16.8$\times$, 20.2$\times$, and 20.6$\times$ for 100K, 1M, and 10M datasets, respectively.
We conclude that GenASM provides significant performance benefits and energy efficiency over GPU aligners for short \revIII{reads.



\textbf{Hardware Baselines.}} We compare GenASM with two state-of-the-art hardware accelerators for read alignment: GACT \revonur{(from Darwin~\cite{turakhia2018darwin})} and SillaX \revonur{(from GenAx~\cite{fujiki2018genax})}.

Darwin is a hardware accelerator designed for \emph{long} read alignment~\cite{turakhia2018darwin}. Darwin contains components that accelerate both the filtering (D-SOFT) and alignment (GACT) steps of read mapping. 
The open-source RTL code available for the GACT accelerator of Darwin allows us to estimate the throughput, area and power consumption of GACT and compare it with GenASM for read alignment. In Darwin, GACT logic and the associated 128KB SRAM are responsible for filling the dynamic programming matrix, generating the traceback pointers and finding the maximum score. 
Thus, we believe that it is fair to compare the power consumption and the area of the GACT logic and GenASM logic, along with their associated SRAMs.

In order to have an iso-bandwidth comparison with Darwin's GACT, we compare \revIII{only} a single array of GACT and a single set of GenASM-DC and GenASM-TB, because (1)~GenASM utilizes the high \revonur{memory} bandwidth that PIM provides \revonur{only} to parallelize many sets of GenASM-DC and GenASM-TB, and a single set of GenASM-DC and GenASM-TB does \emph{not} require high bandwidth, and (2)~all internal data of both GenASM and Darwin \revIII{is} provided by local SRAMs. We synthesize both designs \revonur{(i.e., GenASM and GACT)} at \revonur{an iso-PVT (process, voltage, temperature) corner}, with \revonur{the} same number of PEs, and with their optimum parameters.

\begin{figure}[h!]
\centering
\vspace{-2pt}
\includegraphics[width=\columnwidth,keepaspectratio]{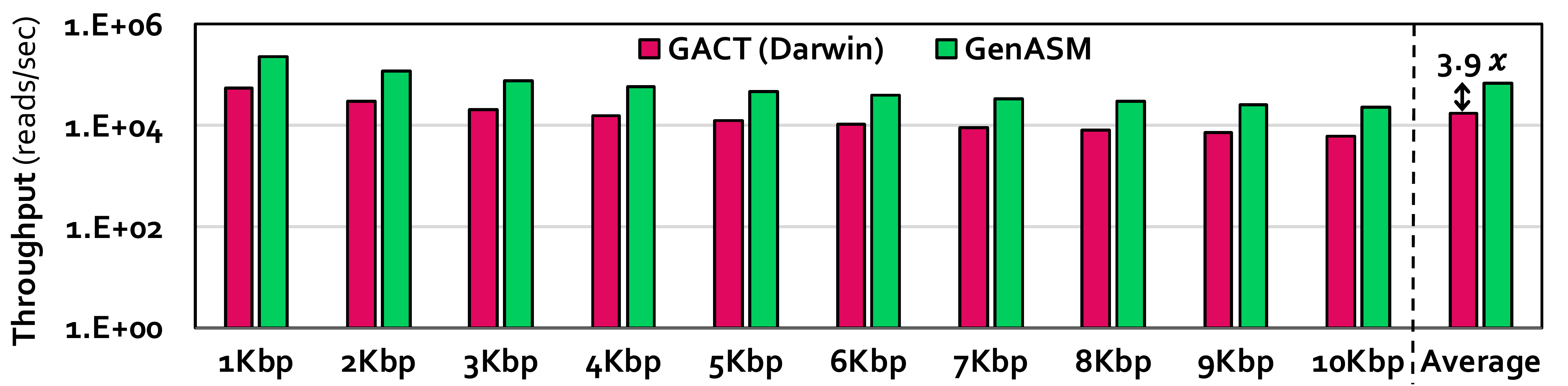}
\vspace{-15pt}
\caption{Throughput comparison of GenASM and GACT \revonur{from} Darwin for long reads.} \label{fig:darwin-long}
\vspace{0pt}
\end{figure}

\revII{As Figure~\ref{fig:darwin-long} shows}, for a single GACT array with 64 PEs at 1GHz, the throughput of GACT decreases from 55,556 to 6,289 alignments per second when the sequence length increases from 1Kbp to 10Kbp, while consuming \SI{277.7}{\milli\watt} of power. In comparison, 
\revIV{for} a single GenASM accelerator at 1GHz (with a 64-PE configuration), the throughput decreases from \hl{236,686} to \hl{23,669} alignments per second when the sequence length increases from 1Kbp to 10Kbp, while consuming \revonur{\SI{101}{\milli\watt}} of power. \revonur{This shows that, on average,} GenASM \revII{provides $3.9\times$} better throughput than GACT, while \revIII{consuming} $2.7\times$ \revIII{less power}. Thus, GenASM provides \revII{$10.5\times$} better throughput per unit power \revonur{for long reads} when compared to GACT.

\begin{figure}[h!]
\centering
\vspace{-2pt}
\includegraphics[width=\columnwidth,keepaspectratio]{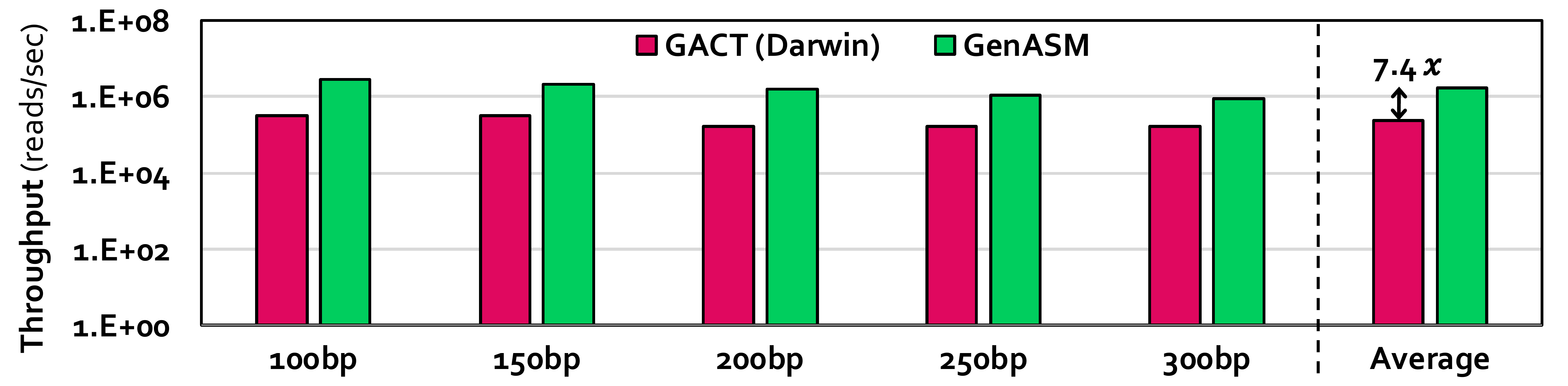}
\vspace{-15pt}
\caption{Throughput comparison of GenASM and GACT \revonur{from} Darwin for short reads.} \label{fig:darwin-short}
\vspace{0pt}
\end{figure}

\revII{As Figure~\ref{fig:darwin-short} shows,} we also compare the throughput of GenASM and GACT for short read alignment \revII{(i.e., 100--300bp reads)}. We find that GenASM performs \revII{7.4$\times$} better than GACT when 
aligning short reads, \revII{on average}. \revonur{Thus, GenASM provides \revII{$20.0\times$} better throughput per unit power for short reads when compared to GACT.}

We compare the required area for the GACT logic with 128KB of SRAM and the required area for the GenASM logic (GenASM-DC and GenASM-TB) with 8KB of DC-SRAM and 96KB of TB-SRAMs, at 28nm. We find that \revonur{GenASM requires $1.7\times$ less area than 
GACT. Thus, GenASM provides \revII{$6.6\times$ and $12.6\times$} better throughput per unit area for long reads and for short reads, respectively, when compared to GACT.}

The main difference between GenASM and GACT is the underlying algorithms. GenASM uses \revonur{our} modified 
Bitap algorithm, which requires only simple and fast bitwise operations. On the other hand, GACT uses the complex and computationally more expensive \revV{dynamic programming based algorithm} for alignment. This is the main reason why GenASM is more efficient than GACT of Darwin. 
\revonur{GenAx} is a hardware accelerator designed for \emph{short} read alignment~\cite{fujiki2018genax}. Similar to Darwin, \revonur{GenAx accelerates} both the filtering and alignment steps of read mapping. Unlike GenAx, whose design is optimized only for short reads, GenASM is more robust and works with \emph{both} short and long reads.
While we are unable to reimplement GenAx, the throughput analysis of SillaX \revonur{(the alignment accelerator of GenAx)} provided by the original work~\cite{fujiki2018genax} allows us to \revonur{provide a} performance comparison between GenASM and SillaX for short read alignment. 

We compare SillaX with GenASM at their optimal operating frequencies (2GHz for SillaX, 1GHz for GenASM), and find that GenASM provides $1.9\times$ higher throughput for \revIII{short 
reads} (101bp) than SillaX (whose approximate throughput is 50M alignments per second). 
\revII{Using} the area and power numbers \revII{reported} for the \revIII{computation} logic of SillaX, we find that GenASM requires 63\% less logic area (\SI{2.08}{\milli\meter\squared} vs.\ \SI{5.64}{\milli\meter\squared}) and 82\% less logic power \revIII{(\SI{1.18}{\watt} vs.\ \SI{6.6}{\watt}). 

In} order to compare the total area of SillaX and GenASM, we perform a CACTI-based analysis~\cite{wilton1996cacti} for \revonur{the SillaX SRAM (\SI{2.02}{\mega\byte})}. \revonur{We find that the SillaX SRAM consumes an area of} \SI{3.47}{\milli\meter\squared}, \revonur{resulting in a total area of \SI{9.11}{\milli\meter\squared} for SillaX}. Although GenASM \revonur{(\SI{10.69}{\milli\meter\squared})} requires \revonur{17\% more total area than SillaX,} 
we find that GenASM provides $1.6\times$ better throughput per unit area for short reads than \revonur{SillaX}.

\textbf{Accuracy Analysis.} \label{sec:results-accuracy}
We compare the traceback outputs of GenASM and \revIII{(1)~BWA-MEM} for short reads, \revIII{(2)~Minimap2} for long reads, to assess the accuracy and correctness of GenASM-TB. We find that the optimum $(W,O)$ setting (i.e., window size and overlap size) for the GenASM-TB algorithm in terms of performance and accuracy is $W=64$ and $O=24$. With this setting, GenASM \revIII{completes the alignment of all reads in each dataset, and increasing} the window size does \emph{not} change the alignment output.

For short reads, we use the default scoring setting of BWA-MEM (i.e., match=+1, substitution=-4, gap opening=-6, and gap extension=-1). 
\revonur{For 96.6\% of the short reads, GenASM \revIV{finds} an alignment whose score is equal to the score of the alignment reported by BWA-MEM. This \revV{fraction increases to 99.7\% 
when we consider scores that are within $\pm4.5\%$ of the \revIV{scores reported} by BWA-MEM.}}

For long reads, we use the default scoring setting of Minimap2 (i.e., match=+2, substitution=-4, gap opening=-4, and gap extension=-2). 
\revonur{For 99.6\% of the long reads with a 10\% error rate, GenASM \revV{finds} an alignment whose score is within $\pm0.4\%$ of the score of the alignment reported by Minimap2. For 99.7\% of the long reads with a 15\% error rate, GenASM \revV{finds} an alignment whose score is within $\pm0.7\%$ of the score of the alignment reported by Minimap2.}

\revmicro{
There are two reasons for the \revonur{difference between the alignment scores reported by GenASM and the scores reported by the baseline tools.} First, GenASM performs \revonur{traceback} for the alignment with the minimum edit distance. However, the \revonur{baseline can report an alignment that has a higher number of edits but a lower score than the alignment reported by GenASM}, when more complex scoring schemes are used. Second, during the TB stage, GenASM follows a fixed order at each iteration when picking between substitutions, insertions, or deletions (based on the penalty of each error type). \revonur{While} we pick the error type with the lowest possible cost at the current iteration, another error type with a higher \revonur{initial} cost may lead to a better (i.e., lower-cost) alignment in later iterations, which cannot be known beforehand.\footnote{\revIII{We can add support for different orderings by adding more configurability to the GenASM-TB accelerator, which we leave for future work.}}}

Although GenASM is optimized for \revonur{unit-cost based scoring (i.e., edit distance) and \revIII{currently} provides only \revV{partial} support} for more complex scoring schemes, we show that GenASM framework can still serve as a fast, memory- and power-efficient, \revIII{and quite accurate} alternative for read alignment.


\subsection{Use Case 2: Pre-Alignment Filtering} \label{sec:results-filtering}

\rev{
We compare GenASM with the state-of-the-art FPGA-based pre-alignment filter for short reads, Shouji~\cite{Alser2019}, \revIV{using} two datasets provided in~\cite{Alser2019}. When we compare 
Shouji (with maximum filtering units) and GenASM for the dataset with 100bp sequences, we find that GenASM provides $3.7\times$ speedup over Shouji, \revonur{while reducing power consumption by $1.7\times$}. When we perform the same analysis with 250bp sequences, we find that \revonur{GenASM does not provide speedup over Shouji, but reduces power consumption by $1.6\times$}.

In pre-alignment filtering for short reads, 
\revII{only} GenASM-DC is executed \revIII{(Section~\ref{sec:bitmac-framework-filter})}. The complexity \revonur{of} GenASM-DC 
is \revII{$O(n\times m \times k)$} whereas the complexity of Shouji is \revII{$O(m\times k)$}, where $n$ is the text length, $m$ is the read length, and $k$ is the edit distance threshold. \revIII{Going from the 100bp dataset to the 250bp dataset}, all these three parameters increase linearly. Thus, \revonur{the} speedup of GenASM over Shouji \revonur{for pre-alignment filtering} decreases for datasets with longer reads.

\revII{To analyze filtering accuracy}, \revonur{we use Edlib~\cite{vsovsic2017edlib} to generate the ground truth edit distance value for each sequence pair in the datasets (similar to Shouji)}}. 
We evaluate the accuracy of GenASM as a pre-alignment filter by computing its false accept rate and false reject rate \revonur{(as defined in \cite{Alser2019})}.

The false accept rate\revonur{~\cite{Alser2019}} is the ratio of the number of dissimilar sequences that are falsely accepted by the filter \revIII{(as similar)} and the total number of dissimilar sequences that are rejected by the ground truth. The goal is to minimize the false accept rate to maximize the number of dissimilar sequences that are eliminated \revIII{by the filter}. For the 100bp dataset 
with \revonur{an} edit distance threshold of 5, Shouji has \revonur{a 4\% false accept rate, whereas GenASM has a false accept rate of only 0.02\%}. 
For the 250bp dataset with \revonur{an} edit distance threshold of 15, Shouji has \revonur{a 17\% false accept rate, whereas GenASM has a false accept rate of only 0.002\%.} 
Thus, GenASM provides a very low rate of \revonur{falsely-accepted} dissimilar sequences, and significantly improves the accuracy of pre-alignment filtering compared to Shouji. 

\revonur{While Shouji approximates the edit distance, GenASM calculates the actual distance. Although calculation requires more computation than approximation, a computed distance results \revIII{in} a near-zero (0.002\%) false accept rate.\footnote{\revIV{The reason for the non-zero false accept rate of GenASM is \revIV{that} when there is a deletion in the first character of the query, 
GenASM does \emph{not} count this as an edit, and skips this extra character of the text when computing the edit distance. Since GenASM reports an edit distance that is one lower than the edit distance reported by the ground truth, if GenASM's reported edit distance is below the threshold but the ground truth's is not, GenASM leads to a false accept.}}
Thus, GenASM filters more false-positive locations out, leaving fewer candidate locations for the expensive alignment step to process. This \revIII{greatly} reduces the combined execution time of filtering and alignment. Thus, even though GenASM does not provide any speedup over Shouji when filtering the 250bp sequences, its lower false accept rate makes it a better option for this step of the pipeline with greater overall benefits.} 


The false reject rate\revonur{~\cite{Alser2019}} is the ratio of the number of similar sequences that are rejected by the filter \revIII{(as dissimilar)} and the total number of similar sequences that are accepted by the ground truth. The false reject rate should always be equal to 0\%. We observe that GenASM always \revonur{provides a 0\% false reject rate, and thus does} not filter out similar sequence pairs, \revIII{as does Shouji}.

\subsection{Use Case 3: Edit Distance Calculation}\label{sec:results-edit}

\begin{figure}[h!]
\centering
\vspace{-2pt}
\includegraphics[width=\columnwidth,keepaspectratio]{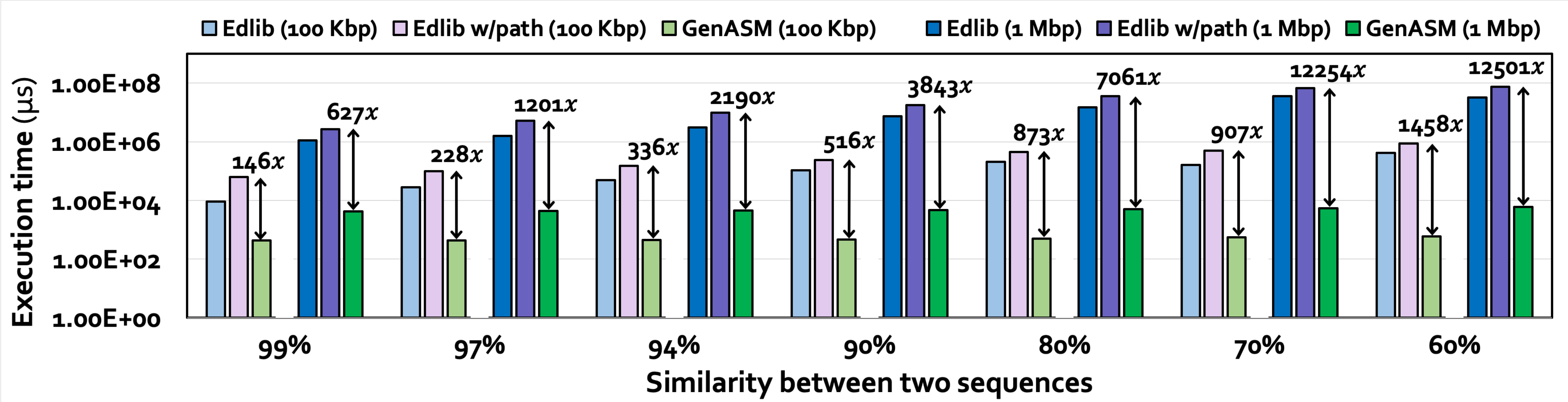}
\vspace{-15pt}
\caption{Execution time comparison of GenASM and Edlib for edit distance calculation.} \label{fig:execution-result-edlib}
\vspace{-2pt}
\end{figure}

We compare GenASM with the state-of-the-art edit distance calculation library, Edlib~\cite{vsovsic2017edlib}. Figure~\ref{fig:execution-result-edlib} compares the execution time of Edlib (with and without finding the traceback \revonur{output}) and GenASM when finding the edit distance between \revonur{two sequences of length 100Kbp, and also two sequences of length 1Mbp}, which have similarity ranging from 60\% to 99\% (Section~\ref{sec:methodology:datasets}). 
\revonur{Since Edlib is a \revIII{single-thread} edit distance calculation tool, for a fair comparison, we compare the throughput of only \revonur{one} GenASM accelerator \revIII{(i.e., in one vault)} with \revIII{a single-thread} execution of the Edlib tool.}

As Figure~\ref{fig:execution-result-edlib} shows, when performing edit distance calculation between two 100Kbp sequences, GenASM \revIII{provides 22--716$\times$ and 146--1458$\times$ speedup over Edlib execution without and with traceback, respectively}. GenASM has the same execution time for both of the cases. When the sequence length increases from 100Kbp to 1Mbp, the execution time of GenASM increases linearly \revIII{(since $W$ is constant, but $m+k$ increases linearly)}. However, due to its quadratic complexity, Edlib cannot scale linearly. Thus, for the edit distance calculation of 1Mbp sequences, GenASM \revIII{provides 262--5413$\times$ and 627--12501$\times$ speedup over Edlib execution without and with traceback, respectively.}

Although both \revonur{the} GenASM algorithm and Edlib’s underlying Myers’ algorithm~\cite{myers1999fast} use bitwise operations only for edit distance calculation and exploit bit-level parallelism, the main \revIII{advantages of the GenASM algorithm come} from \revIV{(1)~the} divide-and-conquer approach we follow for efficient support for longer sequences, \revIII{and \revIV{(2)~our} efficient co-design \revV{of the GenASM algorithm} with the GenASM hardware accelerator}. 

\revonur{Based on our power analysis, we find that
power consumption of Edlib is \SI{55.3}{\watt} and \SI{58.8}{\watt} when finding the edit distance between two 100Kbp sequences and two 1Mbp sequences, respectively. \revII{Thus,} GenASM reduces power consumption by $548\times$ and $582\times$ over Edlib, respectively.}

\revonur{
We also compare GenASM with ASAP~\cite{Banerjee2019}, the state-of-the-art FPGA-based accelerator for edit distance calculation. While we are unable to reimplement ASAP, the execution time and power consumption analysis of ASAP provided in~\cite{Banerjee2019} allows us to provide a comparison between GenASM and ASAP. ASAP is optimized for shorter sequences and reports execution time \revIII{only} for sequences of length 64bp--320bp~\cite{Banerjee2019}. Based on\revV{~\cite{Banerjee2019},} 
the execution time of one ASAP accelerator increases from \SI{6.8}{\micro\second} to \SI{18.8}{\micro\second} when the sequence length increases from 64bp to 320bp, while consuming \SI{6.8}{\watt} of power. In comparison, we report that the execution time of one GenASM accelerator increases from \SI{0.017}{\micro\second} to \SI{2.025}{\micro\second} when the sequence length increases from 64bp to 320bp, while consuming \SI{0.101}{\watt} of power. This shows that GenASM provides 9.3--400$\times$ speedup over ASAP, while \revIII{consuming $67\times$ less power}.
}

\subsection{Sources of Improvement in GenASM} 

\revonur{GenASM's} performance improvements come from our \revonur{algorithm/hardware} co-design, \revIII{i.e.,} both from our modified algorithm and our co-designed architecture for this algorithm. The sources of the large improvements in GenASM are (1)~the very \revIII{simple} computations it performs; (2)~the divide-and-conquer approach we follow, which makes our design efficient for both short and long reads despite their different error profiles; and (3)~the very high degree of parallelism obtained with the help of specialized compute units, dedicated SRAMs for both GenASM-DC and GenASM-TB, and the vault-level parallelism provided by \revonur{processing in the logic layer of 3D-stacked memory}. 

\textbf{Algorithm-Level.} \revII{
Our divide-and-conquer approach allows us to decrease the execution time of GenASM-DC from \revII{$(\frac{m \times(m+k)\times k}{P{\times}w})$} cycles to $((\frac{W{\times}W{\times}min(W,k)}{P{\times}w}){\times}\frac{m+k}{W-O})$ cycles, where $m$ is the pattern size, $k$ is the \revV{edit distance threshold}, $P$ is the number of PEs that GenASM-DC has (i.e., 64), $w$ is the number of bits processed by each PE (i.e., 64), $W$ is the window size (i.e., 64), and $O$ is the overlap size between windows (i.e., 24). Although the total GenASM-TB execution time does \emph{not} change ($(m+k)$ cycles vs. $((W-O)\times\frac{m+k}{W-O})$ cycles),
our divide-and-conquer approach helps us decrease the GenASM-DC execution time by $3662\times$ for long reads, and by $1.6-3.9\times$ for short reads. 
}


\textbf{Hardware-Level.} GenASM-DC's systolic-array-based design removes the data dependency limitation of the underlying Bitap algorithm, and provides $64\times$ parallelism by performing 64 iterations of the GenASM-DC algorithm \revV{in parallel}. 
Our hardware accelerator for GenASM-TB makes use of specialized per-PE TB-SRAMs, which eliminates the otherwise very high \revonur{memory} bandwidth consumption of traceback and enables efficient execution.

\textbf{Technology-Level.} With the help of \revonur{3D-stacked memory's} vault-level parallelism, we can obtain $32\times$ parallelism by performing 32 alignments \revIII{in parallel in} different vaults.

\vspace{-5pt}
\section{\rev{Other Use Cases of GenASM}}\label{sec:bitmac-framework-other}

\revIII{We have quantitatively evaluated three use cases of approximate string matching for genome sequence analysis (Section~\ref{sec:results}). We discuss \revonur{four} other potential use cases of GenASM, whose evaluation we leave for future work.}

\textbf{Read\revIV{-to-Read Overlap} \revIII{Finding Step of de} Novo Assembly.}
\sg{\emph{De novo} assembly\revonur{~\cite{chaisson2015genetic}} is an alternate genome sequencing approach that assembles an entire DNA sequence without the use of a reference genome.}
The first step of \textit{de novo} assembly is \revIII{to find} read-to-read overlaps since the reference genome does not exist\revIII{~\cite{cali2017nanopore}}.
\revonur{Pairwise read alignment (i.e., read-to-read alignment)} is the last step of read-to-read overlap finding\revonur{~\cite{pevzner2001eulerian,li2018minimap2}}. As sequencing devices can introduce errors to the reads, read alignment in overlap finding also needs to take these errors into account. \sg{GenASM can be used for the \revII{pairwise read} alignment step \revII{of} overlap finding.}

\textbf{Hash-Table Based Indexing.}
In the indexing step of read mapping, the reference genome is indexed and stored as a hash table, \revIII{whose keys are all possible fixed-length substrings (i.e., seeds) and whose values are \revIV{the locations} of these seeds in the reference genome.} 
\revII{This index structure is queried in the seeding step to find \revIV{the} candidate matching 
locations of query reads. As we need to find \revIV{the locations} of each seed in the reference text \revIII{to form the index structure}, GenASM can be \revIV{used 
to} generate \revIV{the hash-table based index}.

\textbf{Whole Genome Alignment.}
Whole genome alignment~\revII{\cite{dewey2019whole, paten2011cactus}} is the method of aligning two genomes (from \sg{the same or different} species) for predicting evolutionary \revIII{or familial} relationships between \revIII{these genomes. In whole genome alignment, we need to align two very long sequences. Since GenASM can operate on arbitrary-length sequences \revIV{as a result} of our divide-and-conquer approach, whole genome alignment can be accelerated using the GenASM framework.}}


\revonur{
\textbf{Generic Text Search.} Although GenASM-DC is optimized for genomic sequences (i.e., DNA sequences), which are composed of only 4 characters (i.e., A, C, G and T), GenASM-DC can be extended to support larger alphabets, thus enabling generic text search. \revV{When generating the pattern bitmasks during the pre-processing step, the only change that is required is to generate bitmasks for the \revIII{entire} alphabet, instead of for only four characters.} There is no change required to the edit distance calculation step.

As special cases of general text search, the alphabet can be defined as RNA bases (i.e., A, C, G, U) for RNA sequences or as amino acids (i.e., A, R, N, D, C, Q, E, G, H, I, L, K, M, F, P, S, T, W, Y, V) for protein sequences. This enables GenASM to be used \revIII{for 
RNA} sequence alignment or protein sequence alignment~\cite{haque2009pairwise,smith1981identification,needleman1970general,lipman1985rapid,altschul1990basic,altschul1997gapped,kent2002blat,needleman1970general,notredame2000t,higgins1988clustal,thompson1994clustalw,edgar2004muscle,zou2015halign}. 

}


\vspace{-5pt}
\section{Related Work} \label{sec:related-work}

To our knowledge, this is the first approximate string matching \revIII{acceleration} framework that enhances and accelerates the Bitap algorithm, and \revIII{demonstrates the effectiveness of the framework for multiple use cases in genome sequence analysis.}
Many previous works have attempted to improve (in software or in hardware) the performance of a \emph{single} step of the genome sequence analysis pipeline. Recent acceleration works tend to follow one of two key \revIII{directions\revonur{~\cite{alser2020accelerating}}.



The} first approach is to build pre-alignment filters that use heuristics to first check the differences between two genomic sequences before using the computationally-expensive approximate string matching algorithms. 
Examples of such filters are the Adjacency Filter \cite{Xin2013} that is implemented for standard CPUs, SHD \cite{Xin2015} that uses SIMD-capable CPUs, and GRIM-Filter \cite{Kim2018} that is built in 3D-stacked memory. Many works also exploit the large amounts of parallelism offered by FPGA architectures for pre-alignment filtering,
such as GateKeeper \cite{gatekeeper}, MAGNET \cite{Alser2017}, Shouji \cite{Alser2019}, and \revonur{SneakySnake~\cite{alser2019sneakysnake}}. 
A recent work, GenCache~\cite{nag2019gencache}, proposes an in-cache accelerator to improve the filtering (i.e., seeding) mechanism of GenAx (for short reads) by using in-cache operations~\cite{aga2017compute} and software modifications.

The second approach is to use hardware accelerators for the computationally-expensive read alignment step. 
\revonur{Examples of such hardware accelerators are RADAR~\cite{huangfu2018radar}, FindeR~\cite{zokaee2019finder}, and AligneR~\cite{zokaee2018aligner}, which make use of ReRAM based designs for faster FM-index search, or RAPID~\cite{gupta2019rapid} and BioSEAL~\cite{kaplan2019poster}, which target dynamic programming acceleration \revonur{with} \revIII{processing-in-memory}.
\revIII{Other read alignment acceleration works include} SIMD-capable CPUs~\cite{Daily2016}, multicore CPUs~\cite{Georganas2015, Liu2015}, and specialized hardware accelerators such as GPUs (e.g., GSWABE~\cite{Liu2015}, CUDASW++ 3.0~\cite{Liu2013}), FPGAs (e.g., FPGASW~\cite{Fei2018}, ASAP~\cite{Banerjee2019}), or ASICs (e.g., Darwin~\cite{turakhia2018darwin} and GenAx~\cite{fujiki2018genax})}. 

\rev{In contrast to GenASM, all of these prior works focus on accelerating \revonur{only} a single \revV{use case in genome sequence analysis}, whereas GenASM \revonur{is capable of accelerating at least} three different use cases (i.e., read alignment, pre-alignment filtering, edit distance calculation) where approximate string matching is required.}


\vspace{-6pt}
\section{Conclusion} \label{sec:conclusion}
\vspace{-3pt}
We propose GenASM, an approximate string matching \revIV{(ASM)} acceleration framework for genome sequence analysis \hl{built upon} \revonur{our modified and enhanced} Bitap algorithm. GenASM performs bitvector-based 
\revIV{ASM}, which can accelerate multiple steps of genome sequence analysis. 
We co-design our \revonur{highly-parallel,} scalable and memory-efficient algorithms with low-power and area-efficient hardware accelerators. 
We evaluate GenASM for three different use cases of ASM in genome sequence analysis \revIV{for both short and long reads}: read alignment, pre-alignment filtering, and edit distance calculation. 
We show that GenASM is significantly faster and more power- and area-efficient
than state-of-the-art software and hardware tools for each of these use cases.
\revonur{We hope that GenASM inspires future work in co-designing \revIV{algorithms and hardware} together to create \revIII{powerful} frameworks that accelerate other \revIII{bioinformatics \revIV{workloads} and} emerging applications.}


\vspace{-5pt}
\section*{Acknowledgments}
\vspace{-2pt}
Part of this work was completed during Damla Senol Cali's internship at Intel Labs. This work is supported by funding from Intel, the Semiconductor Research Corporation, the National Institutes of Health (NIH)\revonur{, the industrial partners of the SAFARI Research Group}\revII{, and partly by EMBO Installation Grant 2521 awarded to Can Alkan}. We thank the anonymous reviewers of MICRO 2019, ASPLOS 2020, ISCA 2020, and MICRO 2020 for their comments.


\patchcmd{\thebibliography}{\clubpenalty4000}{\clubpenalty10000}{}{}     
\patchcmd{\thebibliography}{\widowpenalty4000}{\widowpenalty10000}{}{}   
\patchcmd{\bibsetup}{\interlinepenalty=5000}{\interlinepenalty=10000}{}{} 

\bibliographystyle{IEEEtranS}
\bibliography{refs}

\end{document}